\documentclass[showpacs,showkeys]{revtex4}
\usepackage[dvips]{graphicx}
\usepackage{epsfig}
\begin{document}

\hspace{4cm}  \vspace{4cm}

\title{Elastic and Proton Dynamics of the DNA}

\author{Voislav Golo}
\email{voislav.golo@gmail.com }

\affiliation{Department of Mechancs and Mathematics \\
         Moscow University \\
         Moscow 119 899, Russia }

\date{March 27, 2008}

\begin{abstract}
    The subject of this report is the  dynamics of  elastic system in conjunction with hydrogen bonds
    of the DNA. We draw attention to the draw-back of the familiar rod model of the DNA, and make a case of
    constructing models that could accommodate the intrinsic structure  of the DNA.  In this respect
    studying the interplay among the elastic system and  the protons of the DNA, is of interest, for
    it could  accommodate the inter-strand   as  well as the tunneling  modes of protons.
    Following this direction, we come to the conclusion that the elastic-proton dynamics may have a bearing
    on biophysics of the DNA.  The phenomenon of point mutations is discussed within this framework.
\end{abstract}


\vspace{1cm}

\maketitle

\vspace{2cm}

\section{\label{sec:intr}   Introduction}

A molecule of the DNA can attain several hundred $\mu m$ in
length.  If we neglect details that have a size of one thousand
$\AA$, or less, we can visualize it as a soft shapeless line and
conclude that on this scale it behaves like an ordinary polymer.
In contrast, looking at its smaller segments, of one hundred $\AA$
or less, we observe that it tends to be straight.  Thus, borrowing
a comparison from everyday life, we may say that a molecule of the
DNA looks like a piece of steel wire whose long segments are
flexible and the short ones are stiff. The elastic properties of
the DNA are intimately related to its being a double helix. The
latter imposes severe constraints on deformations which can be
effected without destroying the molecule and to a large extent
determines its mechanical properties. In fact, the two strands
comprising the molecule of DNA have just small bending rigidities,
just as usual polymers. But the formation of the two-stranded
structure drastically changes the DNA by making it both stiff and
capable of forming sophisticated spatial shapes. Similarly, the
dynamics of the DNA  is largely determined by the relative motion
of its strands. The functioning of the DNA  also involves the
charge transport in the molecule. The latter may be related either
to the motion of electrons along the strands, or to the tunneling
of protons inside the hydrogen bonds connecting the bases. The
problem that still waits its answer is whether there is any
interaction between the elastic dynamics of the double helix and
the tunneling dynamics of charge.

The diameter of the DNA is about $20 \AA$, and taking into account
its enormous length we may consider it both a microscopical and
macroscopical object. Thus, the DNA requires a special means for
analysis of its physical properties. Important, the approach from
'first principles' aimed at conformational dynamics , for example
using the methods of quantum chemistry, often comes across very
serious difficulties, which do not have just technical character,
but correspond to the need for a physical picture to rely upon.
Therefore, the theoretical study of the DNA is to utilize models,
which are necessarily based on extremely crude simplifications.
Their connection with the microscopical structure of the DNA, is a
matter of intuition rather than of rigorous demonstration. Yet it
would not be quite appropriate to accept only rigorously
demonstrated facts and to ignore ideas that could motivate the
development of the DNA theory and provide a stimulus to further
experimentation.

Owing to the large stiffness of the molecule of DNA, described by
the persistence length of about $500 \AA$ at which it strongly
resists strains caused by heat fluctuations, one may try to
visualize it as an elastic rod. The above model serves the basis
for a number of theoretical approaches, but even though it has
turned out to be successful, for example for the topological
analysis of the DNA conformations, there are situations in which
it does not work properly; specifically, when external charges are
to be taken into account. It is also important that it does not
allow for the internal degrees of freedom due to the two-stranded
helical structure. The latter imposes stringent constraints on
possible strains of a molecule of the DNA, which result in the
specific structure of its vibrational modes, very important for
understanding its functioning. Therefore, we shall consider
consequences that may be inferred from the basic properties of the
DNA, aiming at a qualitative approach that uses simple theoretical
models that may accommodate the internal degrees of freedom. At
this point we should like to note that since different problems
require the use of appropriate approximations and specific choice
of dynamical variables, there is no unique form of the energy
functional, even though the object of our study remains a
molecular of the DNA.  Indeed, we shall choose  various forms of
the energy owing to the necessity to accommodate conformations of
the DNA under consideration and external conditions in action.

Let us recall that the double helix of DNA consists of long
chains, or strands, which have the backbones composed of sugar and
phosphate residues, and special chemicals, bases, keeping the two
strands together. The fundamental building blocks of the strands
are nucleotides, joined to each other in polynucleotide chains.
The nucleotide consists of a phosphate joined to a sugar
(2'-deoxyribose), to which a base is attached. The sugar and base
alone are called a nucleoside. The chains, or strands, of the DNA
wind round each other in a spiral forming a double helix, the
bases being arranged in pairs: adenine - thymine (AT), guanine -
cytosine (GC), so that the sequence of bases in one strand
determines the complimentary sequence of bases in the other and
constitutes the genetic code stored by the molecule of  DNA. There
are several forms of the DNA, denoted by A,B, and Z. The most
common one in nature, is the so-called B-form. One turn of the
helix of the B-form, corresponds approximately to $10.5$
base-pairs, and the distance between adjacent pairs of bases is
approximately $3.4 \AA$. In real life there are considerable
deviations from the canonical B-form of the DNA. Therefore, there
is a need for a special nomenclature for describing its
conformations (see \cite{calladine} for the details). In fact, it
is easy to see that generally a considerable set of parameters are
required to this end. For example, even if we assume a simple
picture of base pairs as flat plates, we still need to use vector
quantities for describing their distance from each other and
angles for mutual orientation, that is on the whole nine
parameters. The most common one, Roll and Tilt, shall describe the
planes of successive base pairs  not being parallel, as is
prescribed for the B-form. In fact, it is necessary to introduce
the so-called propeller angle for defining the measure for
deviations of the normals to the bases inside a base-pair. It is
worth noting that the deviations from the canonical form are by no
means small, and may have a size of tens of degrees.

In contrast to the sugar-phosphate backbone of a single strand of
DNA, which is formed by strong covalent forces, the double helix
of the DNA is due to the interplay of {\it weak} chemical forces.
In fact, the two chains of the DNA are held together by the
hydrogen bonds between complementary bases and the stacking
interactions between adjacent bases attached to the same
sugar-phosphate strand above each other in neighbouring pairs, so
that on the whole it requires the energy of order $10 Kcal/mol$ or
several $k_B T$, whereas the covalent bonds are by orders of
magnitude larger, see \cite{Saenger}. The main contribution to the
thermodynamic stability of the double helix is due to the hydrogen
bonds which require the specific choice of the pairs, adenine -
thymine, guanine-cytosine. The important point about it is that
there is a competition between the hydrogen bonds formed by
base-pairs and the hydrogen bonds of the molecule with water in
aqueous solution\label{Hbonds}. It is generally believed, even
though it is worthwhile to mention that there are no adequate
theoretical estimates to the effect (see p.\pageref{HWcomp}, where
we discuss a rough qualitative model of the phenomenon)  that the
competition makes for the increase of entropy, and thus the
stability of the DNA. The stacking interaction is due to the bases
being flat water-insoluble molecules, lying above each other
roughly perpendicular to the direction of the helical axis so as
to enable electron clouds between  bases  to contribute to the
helical stability. This is again only a plausible hypothesis, or
even simply an intuitive picture that needs careful investigating.
The energy of the interaction between complementary base pairs has
been estimated by various means which rely on experimental data
and computer calculations within the framework of the quantum
chemistry, \cite{SantaLucia}. It is generally falls within $10
Kcal/mol$ (see \cite{SantaLucia} for  more recent results).

The large persistence length of the DNA, $\approx 500 \AA$, which is more than $20$ times larger its
diameter, serves the main, and essentially intuitive, argument for considering it as an elastic rod
and employing the methods of continuum mechanics for its study. Strictly speaking, this is not
correct, for the cross section of this 'rod' corresponding just to a base-pair comprises only several
tens of atoms. The main argument in favour of such approach is generally the pragmatic one, "the truth
is useful" . In fact, as follows from numerical simulation within the framework of this approach, the
above model appears to be acceptable for many regimes that involve functioning of the DNA ,
\cite{marko}, \cite{sig1}. It needs some modifications so as to take into account the important
effects of the anisotropy  due to unsymmetrical positions of constituent nucleoside, and thus requires
the use of anisotropic elastic modulii. Equally important, there is also a geometrical asymmetry
generated by the relative positions of  nucleosides inside a base pair. The asymmetry can be
visualized as two groves on the surface of the rod. The  grooves are helpful in describing the
interactions of the DNA with external charges by allowing the graphic representation of their
distribution  on the surface of a molecule of DNA. It should be noted as well  that a molecule of the
DNA is itself negatively charged. Therefore, the emerging picture of the DNA conformation due to
elastic and electrostatic forces is generally very complicated, and should be treated within the
framework of electro-elasticity theory. But the available experimental  values of elastic stretch
constants and the dependence of persistence lengths  on the ionic strength of aqueous solution,
indicate that, strictly speaking, a molecule of the DNA does not behave like an elastic rod,
\cite{bloomfield}. The effect could be due to the backbones of the two DNA strands, which contain
phosphoric groups carrying  negative charge that may cause the strands to repel each other, facilitate
the separation of the strands, and make for the double-helix being less stable, at low ionic strength,
\cite{bloomfield}. The overall picture strongly depends on the specific arrangement of the constituent
base pairs, because the forces keeping a base-pair together depend on the choice of constituent
nucleotides, the number of the hydrogen bonds involved being different, e.g. 2 for adenine - thymine
and 3 for guanine - cytosine. It is also important, that the relative positions of  bases  change from
pair to pair. Obviously, we are very far from the simple elastic rod model. Therefore, there is a need
for a semi-microscopical theory that could accommodate the elasticity of the DNA and take into account
its micro-structure.

The current approach to the problem relies on the hypothesis that
it is possible  to separate the dynamical modes of the bases and
the sugar-phosphate backbone by considering the coupling between
them as perturbation. The backbone modes are suggested to be
strongly overdamped, whereas the  modes due to the motion of the
bases, or  inter-strand modes, are assumed to be less sensitive to
the viscosity of ambient liquid. The theoretical calculations of
paper \cite{georghiou} support the statement, whereas the
experiments on Raman scattering, \cite{urabe2},\cite{urabe},
indicate that the attenuation of the modes is substantial, see
also \cite{proh1} for the theoretical treatment of the
inter-strand modes.\label{attenuation}

The approaches commonly used to manufacture qualitative models
that could give an adequate picture of the dynamics of the DNA,
are generally  based on the concept of lattice ( see paper
\cite{kovaleva1}, \cite{kovaleva2}, \cite{us} in which a
coarse-grained model is considered), that is one considers the
molecule of DNA as a regular structure similar to a
one-dimensional crystal. It is necessary to take into account
that: (1) the DNA comprises the two strands; (2) the strands are
bound together by certain forces determining inter-strand motions;
(3)the relative position of the strands verify the helical
symmetry. But, the DNA is not totally symmetrical structure due to
the choice of base-pairs, which is generally random. This
circumstance results in considerable theoretical complications.
Important, the helical structure supposes the existence of the
preferred local system of coordinates for every base-pair, and
these coordinate systems should change from one base-pair to
another, the neighbouring one. Constructions accommodating the
phenomenon utilize the concept of gauge field, following its
current use in field theory and condensed matter physics. In the
case of the DNA dynamics, it allows for at least qualitative
description of its elastic modes.

Using the lattice has also the advantage of allowing for
tautomeric forms of the base pairs.  There are  of two classes of
the DNA bases, purine (adenine, guanine )and pyrimidine (cytosine,
thymine). Under the ordinary circumstances adenine and cytosine
are in amino form, and only rarely in the imino one, whereas
guanine and thymine prefer the keto form, and rarely the enol one.
Inside the base-pairs the transformation of the tautomeric forms
correspond to the tunneling transitions of protons in the hydrogen
bonds keeping the bases together. It is alleged that the
transitions may result in mutations, \cite{CW1}, \cite{CW2},
\cite{Loew}, \cite{Crick}. The intriguing question is whether
there is an interplay among the elastic properties of the DNA and
the tautomeric transitions.

\section{ The elastic rod model}
\label{sec:elrod}

In this section we shall consider more fully the rod model and its
possible modifications. One may advance the hypothesis that the
conformation of the molecule of DNA could be qualitatively
described by visualizing the latter as an elastic thread that has
the elasticity constant, $\gamma$,  of such a size that heat
fluctuations due to surrounding solvent be small on a scale called
the persistence length. In order to make a rough estimate to the
effect, we may assume that the elastic energy of a molecule be
given by the equation
$$
   E_{elastic} \approx \int\limits_0^L \gamma
               \left( \frac{d \phi}{ds} \right)^2 \, ds
$$
in which $L$ is the length of the molecule, $\phi$ is the
deviation angle of the  vector tangent to its central line. Then
the condition for the persistence length, $L_p$, reads
$$
   E_{elastic} \ge k_B T
$$
If we take $L \approx 500 \AA, \quad T \approx 300$ and $\phi \approx 1$ radian, we obtain $\gamma
\approx 10^{-19} \, erg \cdot cm$, that is the value which is in reasonable agreement with the
experimental data, \cite{bloomfield}, \cite{bryant}. Thus, on the spatial scale between one coil of
the double helix and $500 \div 1000 \AA$, that is $10 \div 150$ base pairs a molecule of the DNA may
preserve its straight form and has the appearance of an elastic rod. This model of the DNA allows to
calculate, at least qualitatively, interwound structures called plectonemic supercoils formed by DNA
molecules (first found by electron microscopy, \cite{vinogr}).

Within the framework of this model, the double helix of DNA is
characterized by three spatial scales: (1) the microscale of order
$3.4 \; \AA$, that is the distance between adjacent base pairs
along the chain; (2) the mesoscale of order $10^3 \; \AA$, or
several persistent lengths; (3) the macroscale of the size of a
molecule of DNA, that is up to several $\mu m$ or more. On the
microscale, the molecule of DNA is formed by the base pairs of
purines and pyramidines linked by hydrogen bonds between the
bases; the whole constitutes a double stranded structure. The DNA
helix considered on the mesoscale, is suggested to    have the
properties of an elastic rod with the torsional and the bending
rigidities of about $10^{-19} erg \cdot cm$, \cite{bloomfield}.
The stretching of the rod is assumed to be small compared with the
bending and twisting, and in many cases may be neglected. On the
macroscale, a molecule of DNA is  flexible, its rigidity does not
influence its shape, and one can consider it as usual polymer. The
whole picture constitutes the so-called worm-like-chain model,
\cite{marko}.

The picture of the DNA considered on the mesoscale can be cast in
a quantitative form using the classic theory of elastic rod worked
out by G. Kirchhoff, \cite{LL}. It should be noted that the use of
the Kirchhoff theory for the needs of the DNA involves certain
approximations. First, there is a problem of taking into account
the finite diameter of the molecule, for strictly speaking the
Kirchhoff equations are written down for the elastic {\it line}.
Second, the use of continuum mechanics for objects with spatial
scales of several tens of $\AA$ may raise some doubts; in fact,
there is no continuous medium at hand. Third, Kirchhoff's model
does  not allow for the possible extension of a molecule of the
DNA, and the internal degrees of freedom, for example the relative
motion of the the strands. Nonetheless, the cautious employment of
the Kirchhoff theory gives reasonable qualitative results and
appears to be justifiable in certain regimes, \cite{marko}.

The static of a system of this kind can be described by means of
the effective energy, which constitutes the core of Kirchhoff's
model. It is given by the equation
\begin{equation}
 \label{ebmom}
    F = \int^L_0 ds \,  \frac{1}{2} \sum_{ij} \, a_{ij} \omega_i \omega_j
\end{equation}
in which $L$ being the total length of rod, $s$ the length
parameter, $a_{ij}$ its elastic moduli, and $\omega_i$ coordinates
of a vector that describes the strain of the rod corresponding to
the molecule.  The vector $\vec \omega$ is constructed as follows.
Consider a local frame defined at a point of the central line of
the rod; its first vector $\vec v_1$ being the unit tangent vector
at the point, and the second and the third, $\vec v_2, \ \vec v_3$
unit vectors along the principle directions of the strain. The
three vectors are  considered as columns of the rotation matrix $R
= R(s)$ describing the change of the local frame. The matrix
\begin{equation}
    \omega = R^{-1} \frac{d}{ds} R \label{angvel}
    \label{strain1}
\end{equation}
can be visualized as an angular velocity of the local frame, the
length parameter $s$ playing the part of time, or in the vector
form
\begin{equation}
   \omega  = \sum_i \ f^i \omega_i, \qquad (f^{i})_{jk} = -
   \epsilon_{ijk}
   \label{strain2}
\end{equation}
The minimization equations for the Kirchhoff energy have the same
form as the equations of motion for the top (the so-called
Kirchhoff analogy)
$$
   \frac{d}{dt} \, \vec \omega = \vec \omega \times \vec \mu,
   \qquad
   \vec \mu = \displaystyle{  \frac{\partial F}{\partial \vec \omega}  }
$$
The rich analytical and topological structure of solutions to the
top, or Kirchhoff equations has provided the necessary framework
for their application to  conformational problems of the DNA, and
the most important one has been the  theory of supercoiling,
\cite{marko}. The Kirchhoff model allows for the twist-bend
coupling, \cite{sig1}, which corresponds to the chiral character
of the DNA. In paper \cite{nelson1} the model is further extended
by taking into account the stretch of a molecule of DNA, and
considering the {\it bend-stretch} coupling. In the notations of
papers \cite{nelson1, nelson2, nelson3,  nelson4}, this means that
the elastic energy of the molecule reads
\begin{eqnarray}
    E = \frac{1}{L} \int\limits_0^L \, ds \,
        \left[
          A^{\prime} \Omega_1^2
        \right.                 &+&
                                    A \Omega_2^2 +
          C(\Omega_3 - \omega_0)^2 + B \omega_0^2 \alpha^2
                                     \nonumber  \\
                                &+&
        \left.
          2 D \omega_0 (\Omega_3 - \omega_0) \alpha  +
          2 G (\Omega_3 - \omega_0) \Omega_2 +
          2 K \omega_0  \Omega_2 \alpha
        \right]
                                     \label{nelson_energy}
\end{eqnarray}
where the vector $\Omega_1, \Omega_2, \Omega_3$ defined by
Eqs.(\ref{strain1} - \ref{strain2}) determines the bend and twist
of the molecule, the constant  $\omega_0$, in fact the vector
$\omega_1 = 0, \, \omega_2 = 0, \, \omega_3 = \omega_0$,
accommodates the helix twist, and $\alpha$ accounts for the
stretch. Thus, $C$ is the twist-bending coupling constant of paper
\cite{sig1}, and $K$ the bend-stretch one. The conformations of
the rod is finally determined by the minimum of the functional
given by the equation
\begin{equation}
    {\cal M} = \frac{E}{k_B T} - f Z - 2 \pi \tau  Lk
\label{link}
\end{equation}
in which $f$ is an applied tension and $Lk$ is the topological
invariant, the so-called linking number, that defines the linking
between a closed path and its image obtained by a small
translation in space without self-intersections. The Lagrange
multiplier $\tau$ serves a kind of chemical potential for the
linking. In paper \cite{nelson2} the authors put forward arguments
that there are twist-stretch terms in the elastic energy if the
molecule of DNA is modelled on a stack of thin, rigid plates that
are not permitted to deform. They obtain an equation for the
elastic energy of the form given by Eq.(\ref{nelson_energy}). The
problem is related to that of small fluctuations of bend in the
helical backbone that are important for understanding the
mechanism of the torsional stress accompanying the transcription.
As is shown in paper \cite{nelson3}, the drag could correspond to
a torque of $19^{-13} \, dyn \, cm$, which rather surpasses the
actual torque involved during transcription. In contrast, if the
bend fluctuations are not taken into account, the torque is
negligible, which contradicts the experimental facts. These
results are in qualitative agreement with the value $4.5 \times
10^{-19}$ for the twist rigidity (see  also \cite{lavery}).

In spite of the successes of the elastic rod model, there has been
a considerable criticism levelled at it. Baumann et al,
\cite{bloomfield}, have studied the elastic properties of the DNA
as a function of ionic strength and in the presence of multivalent
cations. They measured the extension of the DNA caused by an
external force applied to it.  It should be noted that there are
three regimes in the elastic response of DNA molecules,
\cite{smith}:
 \begin{enumerate}
    \item $0.01 \div 10 pN$ the molecule behaves as an entropic spring,
          the worm-like chain model (WLC), \cite{khokhlov}; \\
    \item $10 \div 65 pN$  deviations from the WLC, enthalpic elasticity;\\
    \item at about $65 pN$ the molecule suddenly yields in
          a highly cooperative fashion and overstretches
          $\approx 1.7$ times, \cite{smith}.
 \end{enumerate}
In the region where the WLC model is valid, a molecule of the DNA
is a kind of hybrid of a rigid rod and a flexible coil, and is
usually visualized as a homogeneous elastic rod. But the
assumption is in contradiction with the elasticity theory
according to which the persistence length $P$ and the stretch
modulus, $S$, given by the equation $  S =  E A $, in which $E$ is
the Young modulus and $A$ is the cross-sectional area of the rod,
should vary in the same way with the ionic strength whereas the
results of \cite{bloomfield} indicate that they change in opposite
directions. Another discrepancy comes while considering the
Poisson ratio $\sigma$ defined by the equation
$$
     B / C = 1 + \sigma
$$
in which  $B$ and $C$ are the bending and the torsional rigidity,
respectively. The thermodynamical stability requires, \cite{LL},
$-1 < \sigma < 1/2$,  while $\sigma < 0$ corresponds to the
thickening of the rod as it is stretched. The values of $B$ are
alleged to be $2 \cdot 10^{-19} \, erg \cdot cm$,  and $C$ in the
range $2 \cdot 10^{-19}$ to $3.4 \cdot 10^{-19}$,
\cite{bloomfield}. These values correspond to $-0.4 < \sigma <
0.$, and therefore the DNA rod should thicken while it is
stretched. Thus, we should either accept that the DNA is an
elastic rod of quite unusual nature, or admit that there is a need
for a model that could accommodate its double-stranded structure,
helical symmetry, and internal stacked base pairs.

Besides the drawbacks mentioned above the rod model fails to
accommodate the process of denaturation when a molecule of the DNA
splits up into separate strands.  But, the breaking of the
hydrogen bonds between the base pairs and the formation of bubbles
comprising segments in which  the two strands are separated, may
happen under other circumstances. Equally important,  there may
exists  relative motions of the strands which do not result in
breaking the hydrogen bonds. In fact, the dynamics of these
inter-strand modes is accessible to experimental studying, (see
for example \cite{urabe}), \cite{urabe2}, \cite{tao}, \cite{tao2}.
It may tell a lot about the physics of the DNA.

One may try to mimic the partition of a single double-stranded
molecule into two strands within the framework of the rod model,
by employing an additional quantity $\vec q$ that indicates a
relative displacement of the strands from the equilibrium
conformation\label{vectorq}. The procedure is similar to that
discussed above in connection with the {\it coupling between the
stretch and the twist-bend modes of the DNA}.  What's more, we
shall see that thermal fluctuations of the field $\vec q$, result
in a contribution to the twist-bend momentum and thus turn out to
be similar to the stretch term introduced in \cite{nelson1},
\cite{nelson2}.

The field $\vec q(s)$ gives  displacements of points coinciding in
the initial equilibrium conformation of the rod, and possible
going apart because of the deformation and breaking on the
microscopical scale, of bonds between base pairs of the molecule.
On the mesoscale, which we use to describe the molecule, the
vectors $\vec q(s)$ are determined at points of the rod by the
parameter of arc length, $s$. Thus, outside the region of states
where the breaking up takes place, we visualize the molecule of
DNA  as comprised of two elastic wires  attached to each other;
the whole being an elastic rod that one can bend, twist, and
stretch. It is worth noting that on allowing the partition of the
strands we at the same admit that the molecule can be stretched.
In fact, the sugar-phosphate backbones of the strands are formed
by strong covalent bonds and are hard to be stretched, so that the
partitioning of the strands should result in changing the total
length of the total molecular. Whether it will be  diminishing or
increasing, depends on the deformation of  helix  that should
accompany it.

The static of a system of this kind can be described within the
framework of Kirchhoff's model of the elastic line with the help
of its extension with the vector $\vec q$. It is determined by the
energy given by the equation, \cite{gk1}, \cite{gk2},
\begin{equation}
 \label{enpi}
 F = \int^L_0 ds \, \left(
                     \frac{1}{2} \sum_{ij} a_{ij} \omega_i \omega_j
                     + \sum_i b_i \omega_i
                     + \frac{1}{2}A \left[
                           \partial_s \vec q + \vec \omega \times \vec
                           q
                         \right] ^2
                     + U(\vec q)
                   \right)
\end{equation}
in which $L$ is  the total length of rod,   $a_{ij}$ are its elastic modulii, and $\omega_i$
coordinates of the vector of deformations described above for the Kirchhoff model. The potential
$U(\vec q)$ accommodates forces that keep the two strands hanging together. The vector $(b_1, b_2,
b_3)$ is analogous to $\omega_0$ of paper \cite{nelson1}, and describes the molecule's winding
determined by certain external conditions, for example,  a histone.  . We have used the vector $\vec
q$ in the fourth term of equation (\ref{enpi}) so as to accommodate  deformation and breaking of bonds
between base pairs, and as well as the separation of strands. But here, again, it is worthwhile to
note that the term is similar to the fourth term in Eq.(\ref{nelson_energy}), see paper
\cite{nelson1}, Eq.(3); besides the mathematical similarity we should like to draw attention to the
fact that the reason for this lies in the relation of $\vec q$ to stretching the molecule. The third
term in equation (\ref{enpi}) is the covariant derivative
$$
    \nabla_s \vec q = \partial_s \vec q \; + \; \vec \omega \times \vec q
$$
The covariant derivative is in order because we have to consider the displacements of the strands with
respect to the local coordinates determined by the helical structure and the strain described by the
vector $\vec q$.  The procedure is quite common in the theory of gauge fields.

The state of  equilibrium corresponds to the minimum of $F$ and
gives the equations
$$
    \nabla_s^2 \vec q  =  \frac{\partial U}{\partial \vec q} ,  \qquad
    \nabla_s \left( \vec m + A \vec q \times \nabla_s \vec q \right) = 0
$$
in which $\nabla_s$ is the covariant derivative
$$
  \nabla_s \, \vec X =
               \partial_s \vec X + \vec \omega \times \vec X
$$
and $\vec m$ reads
$$
    m_i = \sum_{j=1}^3 a_{ij} \omega_j
$$

Now let us neglect nonlinear, that is greater than  second order
terms, in the potential $U$, and consider the effect of thermal
fluctuations of the field $\vec q(s)$, that is inter-strand
motion, or in the context of paper \cite{nelson1} the stretching,
on the total configuration of the molecule. To that end we need to
average the field $\vec q$ out, and find the effective energy
\begin{equation}
 \label{kardar}
 e^{\displaystyle{ -\beta F_{eff}}} = \int D \vec q \,
                      e^{\displaystyle{ -\beta F}} \ , \quad
                      \beta = 1/kT
\end{equation}
We aim at a specific, but very important, configuration in which
the vector $\vec \omega $ is equal to the constant
\begin{equation}
    \label{vecom0}
        \vec \omega_0 =  - a^{-1} \vec b
\end{equation}
that provides the absolute minimum for the density of Kirchhoff's energy given by Eq.(\ref{ebmom}),
and the conformation of the regular coil for the molecule. Here $a$ is the matrix of Kirchhoff's
modulli $a_{ij}$.

To evaluate the functional integral in Eq.(\ref{kardar}), we shall
employ Feynman's variational principal, \cite{feynman}, and to
that end resolve the expression for the energy~(\ref{enpi}) in the
form $F = F_0 + F_1 $ with $F_0$ given by the equation
$$
 \begin{array}{ll}
    F_0 = \int\limits_0^L \, ds \,
                        & \left(
                                \frac{1}{2} \sum_{ij} a_{ij} \omega_i \omega_j
                                + \sum_i b_i \omega_i
                            \right. \\
                        &        \\
                        &   \left. +  \frac{1}{2} A \, \sum_i \left[
                              {\displaystyle \frac{d}{ds} q_i} \right] ^2 +
                                                \frac{1}{2} \sum_i \left[  B + A \left
                                                [ \omega^2 -  \omega^2_i
                                                \right]
                                                                        \right] q_i^2
                     \right)
 \end{array}
$$
and $F_1$ by the equation
$$
     F_1 = A \int\limits^L_0 \, ds \,
                  \left( \sum_{ijk} \epsilon_{ijk} \omega_i q_j \frac{d}{ds} q_k
                           - \sum_{i \neq j} \omega_i \omega_j q_i q_j
                  \right)
$$
According to Feyman's variational principal,\cite{feynman}, there
is the estimate for free energy $    {\cal F} \leq  {\cal F}_0 +
\,\langle F_1 \rangle_0 $ with $ {\cal F} = F_{fluct}$ of Eq.(
\ref{kardar}), and the other averages being
$$
 e^{\displaystyle{ -\beta {\cal F}_0}} = \int D \vec q e^{{\displaystyle -\beta F_0}}
    \qquad \mbox{and }  \qquad
        \langle F_1 \rangle_0 = \frac{ \int D  \vec q  F_1  e^{{\displaystyle -\beta F_0}}}
                              { \int D \vec q e^{{\displaystyle -\beta F_0}}}
$$
The average $ \langle F_1 \rangle_0 $ gives zero contribution
owing to the Gaussian nature of the integration. Hence, within the
limits of accuracy provided by Feynman's principle, we have $
F_{fluct} = {\cal F} $ and
$$
     F_{eff} =  \int^L_0 ds \
                \left(
                     \frac{1}{2} \sum_{ij} a_{ij} \omega_i \omega_j
                     + \sum_i b_i \omega_i
                \right)
          - kT \sum_i \ln \frac{z_i}{sh(z_i)}
$$
with $z_i$ being given by
$$
 z_i  =  \frac{1}{2}\,L\,\sqrt { \frac{B}{A} + \omega_0^2 - \omega^2_{0i} },
 \quad  i=1,2,3
$$
Here $B/A$ evaluates the coupling between the strands of a
molecule, and $\omega_{0i}$ are coordinates of the vector given by
Eq.(\ref{vecom0}) determining the configuration without
fluctuation corrections.

Let us consider the small coiling and coupling of strands, that is $\omega \ll 1$ and $B/A \ll 1$, or
$z_i \ll 1, i=1,2,3$. Assuming $\vec \omega$ to be constant, we may cast the equation for the
effective energy in the form
$$
     F_{eff}  =  L  \left(
                     \frac{1}{2} \sum_{ij} \left[ a_{ij}  - \frac{1}{3} k T L
                                                \delta_{ij}  \right]
                                                 \omega_i \omega_j
                                                  + \sum_i b_i \omega_i
                                    \right)
    - \frac{1}{24} k T L^2  \frac{B}{A}
$$
By minimizing $F_{eff}$ with respect to $\vec \omega$, we obtain
the correction of the value of $\vec \omega_0$ effecting the
equilibrium conformation. Since the contribution of fluctuations
is assumed to be small, we may write down the corrections to $\vec
\omega_0$ given by Eq.(\ref{vecom0})

\begin{equation}
    \label{fluctb}
    \vec \omega_0^{\displaystyle{fluct}}  =  \vec \omega_0 + \delta \vec \omega_0
    \qquad \mbox{and} \qquad
    \delta \vec \omega_0  =  -  \frac{1}{3} k T L \,
                                                    a^{-2} \cdot \vec b
\end{equation}

From the last equation, we infer that fluctuations of the field
$\vec q$ make for an increase in the coiling of a molecule. It is
easy to estimate a spatial scale on which the arguments given
above are valid; the fluctuation energy should be smaller than the
elastic one, that is counting by orders of magnitudes $ L a
\omega^2 \gg L^2 k T \omega^2 $ which amounts to $ a/L \gg kT $.
For the rigidity of order $10^{-19}\; erg\,cm$ and room
temperatures, one obtains the scale of order $5\;10^{-6}\;cm$,
that is the persistence length. To understand the estimates given
above, let us notice that in the opposite regime, far from
equilibrium, we may set $B \gg 1$, and $B/A \gg \omega^2$. It is
easy to see that in this case there are no corrections of the
values of $\vec \omega$, and consequently no additional increase
in the supercoiling of molecule.

Working within the framework of paper \cite{nelson1}, Moroz and
Nelson found the renormalization of twist stiffness  by bend
fluctuations; they averaged the partition function corresponding
to the functional given by Eq.(\ref{nelson_energy}), obtained the
Schr\"odinger-like equation for the correlator of the tangent
vector to the molecular axis, \cite{nelson4}, \cite{nelson5}, and
found the torque, $\tau(f, \sigma)$, as a function of applied
stress $f$ and external twist $\sigma = \Delta Lk$
\begin{equation}
  \tau(f, \sigma)  =  \frac{\omega_0  \sigma}
          { \displaystyle{ C^{-1} +
                  \left( 4 A \sqrt{A f / k_B T}
                  \right)^{-1}
          }
        }
        \label{nelmoroz}
\end{equation}
Since the setting of the two problems is not identical, we are not
in a position to compare Eqs.(\ref{fluctb}) and (\ref{nelmoroz});
qualitatively, they are related to the same phenomenon of
renormalization of stiffness by bend fluctuations.

It is worth noticing that the hydrogen bonds of DNA can be
deformed and even broken due to the local action of an external
agent, for example enzyme. We may try to describe the process by
an energy term that be included in Eq.(\ref{ebmom})
\begin{equation}
 \label{enz}
    F_{int} = \int\limits^L_0 \, ds \, \left( \vec \beta \cdot \vec q +
                    \vec q \hat \gamma \vec q
                                        \right)
                                \delta \left( s - s_0 \right)
\end{equation}
with $\vec \beta$ and $\hat \gamma$ being a constant vector and a
matrix, respectively. The $\delta$-function factor is used for
describing point-like action of the external disturbance. Consider
the case of the initial value for $\vec b$ equal to zero before
the breaking up of a molecule. For small $\vec q$, using the
linear approximation, we have the following equation for $\vec q$
\begin{equation}
 \label{pi}
    \frac{\partial^2}{\displaystyle \partial s ^2} \vec q - \Omega^2 \vec q =
                 \frac{\vec \beta}{A} \, \delta \left( s - s_0 \right) +
                 \frac{1}{A} \hat \gamma \vec q \delta \left( s - s_0
                 \right)
\end{equation}
with $\Omega^2 = B/A$ . Up to the second order terms the
minimization equation for $\vec \omega$ reads
$$
     a \,   \nabla \vec \omega - A \,
            \nabla
               \left( \frac{d}{ds} \vec q \times \vec q
               \right) = 0
$$
From the last equation one can infer that the external action due
to  Eq.(\ref{enz}) results in the formation of an effective
rotating moment $\vec b_{ext}$ that reads
$$
 \vec b_{ext} = - \frac{G^2 \left( s, s_0 \right)}
                       {A^2} \frac{d}{ds} G
                       \left( s, s_0 \right)
                       \left( \vec \beta \times \hat \gamma
                              \vec \beta
                       \right)
$$
with $G \left( s, s_0 \right)$ being the Green function of Eq.(
\ref{pi}).

From the equation for $G(s,s_0)$ it is easy to see that the size
of the defect changes from $L$, or the mesoscopic scale, for small
$\Omega$, to $1/ \Omega $ for large $\Omega$.

\begin{eqnarray*}
 G \left( s, s_0 \right) & = & - \frac{1}{\displaystyle \Omega sh(\Omega L)}
                            \left(
                                 \theta (s_0 - s) sh(\Omega s) sh \left( (L - s_0)
                                 \Omega
                                 \right)
                            \right.     \\
                                    & + &
                            \left.
                                 \theta (s - s_0) sh(\Omega s_0) sh \left( (L - s) \Omega
                                 \right)
                            \right)
\end{eqnarray*}
with $\vec \beta$ and $\hat \gamma$ being a constant vector and a
matrix, respectively.

The local influence due to a chemical-biological agent may result in the strands of a molecule being
split up on a scale of order $1/ \Omega$ determined by the potential describing the hydrogen bonds, so
there is a kind of Lindemann's criterion: the breaking up of a molecule takes place if $ \sqrt{B/A}
\propto 1/L$. The magnitude of the splitting is determined by the properties of the agent. A drawback
of the model which is based on a quadratic Lagrangian, is that it does not allow for a threshold
effect, which might be taken into account by considering high order terms. In the region of
conformations far from the splitting of a molecule the fluctuations of the field $\vec q(s)$ are
suppressed by the bond due to $U(q)$, and there is no enhancing the supercoiling, in contrast to the
region close to the splitting, where the potential $U(q)$ is effectively small. Since the minimization
of  energy makes sense only on the mesoscale of several persistence lengths, the conformation of a
molecule being determined by entropy on the macroscale , different segments of a molecule should be
broken up independently from each other so as to give a blurred character to the transition
corresponding to the splitting up of the molecule.

There are various arguments to the effect that the double-stranded
DNA should suffer local denaturation and open up locally so that
hydrogen bonds between  base pairs be broken. Besides
physiological processes that involve the 'unzipping', it can take
place spontaneously due to fluctuations owing to the small energy
required, less than $3 k_B T$, \cite{SantaLucia}. Therefore, the
{\it breathing} of DNA is a phenomena that could happen in many
situations.

It was G.S.Manning, \cite{manning}, who suggested that the real
flexibility in the DNA molecule could be due to opening of base
pairs and  the breathing fluctuations be related to elastic
properties of the DNA. The solution to the problem requires the
knowledge of the rate of the breathing fluctuation, and at this
point there is some controversy.

By now there are two methods for studying the breathing
fluctuations: the NMR and the fluorescence correlation
spectroscopy.

The NMR measures the exchange of protons from imino groups with
water, which are suggested only to occur from open base pairs,
\cite{Leroy1}. According to paper \cite{Leroy2} the opening of
base pairs may require prior unwinding or bending of the DNA
double helix, which does not necessarily lead to imino proton
exchange. The life times of base pairs and open states depend on
temperature and  bases involved. In the B-DNA at $15^{o} C$
typical lifetimes range $0.5$ through $7 \, ms$ for $A \cdot T$
pairs and $7$ through $40 \, ms$ for $G \cdot C$. The open state
life time is $10 - 100 ns$, \cite{Leroy2}. Leroy et al,
\cite{Leroy2}, estimate the activation enthalpy for $C4 \cdot G5$
opening to within $45 \div 56 KJ/mol$, or $10.7 \div 13.3
Kcal/mol$ . It is important that the NMR of imino proton exchange
measures the lifetime of a {\it single} base pair, \cite{Leroy3}.

The fluorescence spectroscopy relies on the translation of base
pair fluctuations in fluorescence fluctuations. Altan-Bonnet et
al, \cite{Bonnet}, use synthetic DNA samples containing modified
bases tagged with a fluorophore and a quencher.  When the DNA
structure is closed, the fluorophore and the quencher are in close
proximity and the fluorescence is quenched, it is again restored
if the structure opens so that the fluorophore and the quencher
are pulled apart. Thus, the fluorescence spectroscopy detects the
local denaturation of the DNA, or bubbles of 2 to 10 base pairs
with lifetimes in the $50 \mu s$ range at $37^{o} C$.

The divergence between the NMR and the fluorescence spectroscopy
measurements could be explained by several reasons. First, the NMR
picks up very fast modes owing to the imino proton exchange being
very sensitive to the conformation of a base pair, whereas the
fluorescence spectroscopy picks up large scale deformations of the
DNA. Second, the NMR measures the life time of a formed base pair;
the fluoresceence spectroscopy measures the life time of the open
state.Third, the range of temperatures and the composition of the
strands in the two experimental settings is different. For these
arguments I am indebted to G.Altan-Bonnet.

Thus, the available experimental data indicates that partial
denaturation of the DNA is  a common phenomenon, and it should be
taken into account while using the elastic rod model. The latter
requires, generally, serious modifications when the internal
motions of the double helix need accommodating.

There are arguments of  qualitative nature in favour of  the fact
that the fluctuations of open hydrogen bonds result in an
effective interaction between the strands of the double helix (cf.
p.\pageref{Hbonds} about the bonds between base pairs and those
with water). We shall  describe a hydrogen bond as two-level
system, $|0\rangle$  and $|1\rangle$, with the energy difference
$\epsilon = E_0 - E_1$, and the Hamiltonian $\epsilon \sigma_3$
where $\sigma_3$ is the third Pauli matrix
$$
            \sigma_3 = \frac{1}{2} \ \left[
                                         \begin{array}{cc}
                                             1 & 0 \\
                                             0 & -1
                                         \end{array}
                               \right]
$$
To describe the coupling between the Pauli operators of  hydrogen
bonds between base pairs and the classical system due to the
elasticity of the molecule, we shall employ the field $\vec q$
(see p.\pageref{vectorq}). We suggest that the interaction between
the elastic forces and a single hydrogen bond is small, and
therefore perturbation theory could be employed, that is a kind of
Born-Oppenheimer approximation in the sense that one part of the
system is considered to be classical, whereas the other one
quantum. We cast the equation for energy in the
form\label{heis}
\begin{equation}
    E = F(\omega, \vec q) +  E_{int}(\vec q, \vec \sigma)  + E_{exc}
    \label{qsenergy}
\end{equation}
where $F$ is the energy of elastic rod given by Eq.(\ref{enpi}),
and $E_{exc}$ is the energy of hydrogen bonds,
$$
    E_{exc}  =  - \epsilon \sum\limits_{n=0}^{N-1} \, \sigma_n^3 +
                \beta \sum\limits_{n=0}^{N-1} \,
                \left[ \sigma_n^{+}  \sigma_{n+1}^{-}  +  \sigma_n^{-} \sigma_{n+1}^{+}
                \right]
$$
with $\sigma_n^{\pm}$ being the matrices
$$
        \sigma_n^{\pm} = \frac{1}{2} (\sigma_n^1  \pm  i \sigma_n^2)
$$
The second sum in the above equation accommodates the possible
propagation of opened base pairs. The third term in
Eq.(\ref{qsenergy}) is the interaction between the classical
elastic system and the quantum one given by the hydrogen bonds.
The field $\vec q$ has the sense of a mean field describing the
partition of the strands and its characteristic spatial scale is
hundreds of $\AA$.  The spatial scale of the Pauli operators is a
few $\AA$.

It is important that the operators $\sigma_n^3, \,
\sigma_n^{+},\sigma_n^{-}$ are intimately related to the
conformation of the molecule.

In fact, the breakdown of a hydrogen bond follows a certain
direction in space.  At each site $n$ corresponding to a base pair
of the molecule, there is a local frame formed by unit orthogonal
vectors $ \vec w_1, \, \vec w_2 \, \vec w_3$, for which the vector
$\vec w_1$ is tangent to the axis of the double helix, $\vec w_2$
is normal to the axis, and $\vec w_3$ indicates the direction of
the bond's breakdown. The operators $\sigma_n^3, \,
\sigma_n^{+},\sigma_n^{-}$ indicated above are chosen in accord
with the frame $ \vec w_1, \, \vec w_2 \, \vec w_3$.  Thus, the
term $ \sigma_n^{+}  \sigma_{n+1}^{-}$ in the  energy $E_{exc}$
involves the operators that act in spaces  $(|0\rangle_n, \,
|1\rangle_n)$ and $(|0\rangle_{n+1},  \, |1\rangle_{n+1})$. To
form the interaction term we need to cast $\vec q$  and the Pauli
operators in the same co-ordinate system. By using the unitary
transformation
$$
    s_n^i = U^{-1} \sigma_n^i U
$$
we may find the operators $s_n^i$, and specifically $s_n^3$, that
correspond to $\sigma_n^i$ in the laboratory co-ordinate system
and describe the state of the hydrogen bond at site $n$ from the
point of view of an external observer. In fact, there is the
equation
$$
     U^{-1} \sigma_n^i U   = \sum\limits_{k=1}^3 \, R_{ik}
     \sigma_n^k
$$
in which the matrix $R_{ik}$ corresponds to the rotation that
brings the frame $ \vec w_1, \, \vec w_2 \, \vec w_3$ in the
standard laboratory one.  Therefore, we choose the interaction
energy in the "minimal" form
\begin{equation}
    E_{int}(\vec q, \vec \sigma) = \gamma \,
           \sum\limits_{n=0}^{N-1} \, \vec q_n \, \cdot R_n \,
           \vec \sigma_n
    \label{inter}
\end{equation}
We shall confine ourself to the circular conformation of the
molecule for which the matrices $R_n$ have the form
$$
   R_n = \left(  \begin{array}{ccc}
                   1  &             0    &             0   \\
                   0  &   \cos \phi_n    &  - \sin \phi_n  \\
                   0  &   \sin \phi_n    &    \cos \phi_n  \\
                 \end{array}
         \right)
$$
Therefore the interaction terms read
$$
   \vec q_n \cdot R \vec \sigma_n =
     \left(   \cos \phi_n q^2_n + \sin \phi_n q^3_n
     \right)  \sigma^2_n  +
     \left( - \sin \phi_n q^2_n + \cos \phi_n q^3_n
     \right)
$$
and we have
$$
 \begin{array}{ll}
     E_I + E_{exc} = & \gamma \sum\limits_{n=0}^{N-1}
                         \left[
                               \left( \cos \phi_n q^2_n + \sin \phi_n q^3_n
                               \right) \sigma^2_n  +
                               \left( - \sin \phi_n q^2_n + \cos \phi_n q^3_n
                               \right)
                         \right] \\
                     & - \epsilon \sum_{n=0}^{N-1} \sigma^3_n +
                       \beta \sum_{n=0}^{N-1}
                         \left( \sigma^{+}_n \sigma^{-}_{n+1} +
                                \sigma^{-}_n \sigma^{+}_{n+1}
                         \right)
 \end{array}
$$

Since we are considering the molecule of DNA at temperature far
from denaturation, and therefore in accord with the results of
papers \cite{Leroy1, Leroy2, Leroy3} the number of excitations,
that is of broken hydrogen bonds, is small.  Consequently, we may
utilize the method of approximate secondary quantization,
\cite{tjablikov}, familiar in the theory of spin systems. Let us
consider the Bose operators $b_k, \, b_k^{+}$
$$
    [b_k, \, b_m]  = 0, \quad
    [b_k^{+}, \, b_m^{+}] = 0, \quad
    [b_k^, \, b_m^{+}] = \delta_{km}
$$
By using the substitution
$$
    \sigma_k^3  =  \frac{1}{2}  - b_k^{+} \, b_k, \quad
    \sigma_k^{-} = b_k^{+}, \quad,
    \sigma_k^{+}b_k     =   b_k
$$
we may cast the energy $E_{int}  +  E_{exc}$ in the form

$$
 \begin{array}{l}
   E_I + E_{exc} =  \displaystyle{      - \frac{\epsilon N}{2}
                          +  \frac{\gamma}{2} \sum_{n=0}^{N-1}
                          \left( -\sin \phi_n q^2_n  + \cos \phi_n q^3_n
                          \right)                                    } \\
        \displaystyle{ +  \sum_{n=0}^{N-1}
                \left[
                   - \gamma \left( -\sin \phi_n q^2_n + \cos \phi_n q^3_n  \right)
                   + \epsilon
                \right] b^{+}_n b_n
                   + \beta \sum_{n=0}^{N-1}
                \left( b_n b^+_{n+1} + b_n^+ b_{n+1}
                \right)                                              }  \\
         \displaystyle{ +   \frac{i \, \gamma}{2} \sum_{n=0}^{N-1}
                \left( \cos \phi_n q^2_n  + \sin \phi_n q^3_n
                \right)
                   \left( b^+_n - b_n  \right)                       }
 \end{array}
$$
The first line in the above equation corresponds to the 'vacuum' fluctuations, which may be of
interest as regards the conformation of the double helix given by the matrices $R_n$ and the field
$\vec q$. The above equation  is quadratic with respect to the Pauli operators $b_k, b_k^{+}$ and we
can cast it into the diagonal form by employing Fourier transform
\begin{eqnarray*}
 B_n   & = & \frac{1}{\sqrt{N}} \sum_{n=0}^{N-1}
           e^{\displaystyle \frac{2i \pi  n k}{N}} b_k  \\
 B_n^+ & = & \frac{1}{\sqrt{N}} \sum_{n=0}^{N-1}
           e^{\displaystyle \frac{-2i \pi  n k}{N}} b_k^+
\end{eqnarray*}
so that the energy $E_{int} +  E_{exc}$  be given by the equation
$$
 \begin{array}{l}
    E_I + E_{exc} = const +
        \displaystyle{ \frac{\gamma}{2}\sum_{n=0}^{N-1}
                       \left( - \sin \phi_n \, q^2_n + \cos \phi_n \, q^3_n \right)
        } \\
        +  \displaystyle{ \sum\limits^{N-1}_{k_1 k_2} \epsilon_{k_1 k_2} B_{k_1}^{+} B_{k_2}
              + \sum_{n=0}^{N-1} \left(  A_n B^+_n + A^*_n B_n
                                 \right)
           }
 \end{array}
$$
in which $\epsilon_{k_1 k_2}$  and $A_n$ read
\begin{eqnarray}
 \epsilon_{k_1 k_2} &=& \delta_{k_1 k_2}
                        \left(  \epsilon   + 2 \beta \cos \frac{2 \pi k_1}{N}
                        \right)
                        \label{mainh} \\
                    &-& \displaystyle{ \frac{\gamma}{N}
                            \sum_{n=0}{N-1}
                            \left( - \sin \phi_n q^2_n + \cos \phi_n q^3_n
                            \right)
                                 e^{\displaystyle \frac{ 2 i \pi (k_2 - k_1)n}{N}}
                         };
                         \nonumber    \\
               A_n  &=& \frac{1}{2}
                        \displaystyle{ \frac{i \gamma}{\sqrt{N}}
                              \sum_{n=0}^{N-1}
                                e^{\displaystyle{
                                      \frac{- 2 i \pi (k_2 - k_1)n}{N}}
                                      \left( \cos \phi_n q^2_n + \sin \phi_n q^3_n
                                      \right)
                                   }
                         }
                             \nonumber
\end{eqnarray}
We may eliminate the linear terms in $B_n, \, B_n^{+}$ using the
canonical transformation
$$
        B_n      \rightarrow S^{-1} \, B_n \, S,  \quad
        B_n^{+}  \rightarrow S^{-1} \, B_n^{+} \, S
$$
or more explicitly
$$
    B_n \rightarrow  C_n = B_n + l_n, \quad
    B_n^{+} \rightarrow C_n^{+} = B_n  + l_n^{*}
$$
We obtain the equation
$$
   E_I + E_{exc} = const + E_{conf}
       + \sum\limits^{N-1}_{k_1 k_2} \epsilon_{k_1 k_2} C_{k_1}^+ C_{k_2}
$$
in which the conformation energy $E_{conf}$ reads
 \begin{equation}
 \label{conf}
         E_{conf} =
           \displaystyle{ \frac{\gamma}{2} \sum_{n=0}^{N}-1
              \left( - \sin \phi_n q^2_n + \cos \phi_n q^3_n
              \right)
              + 3 \sum\limits^{N-1}_{k_1 k_2} \left( \epsilon^{-1} \right)_{k_1 k_2}
              A_{k_1}^{*}  A_{k_2}^{*}
          }
\end{equation}
According to the results of Leroy et al \cite{Leroy1, Leroy2,
Leroy3} the effective temperature of the excitations due to the
opening of hydrogen bonds is very low, and therefore the energy of
the system is given by $E_{conf}$.  As was  assumed the
interaction between the elastic part of the total energy and that
due to the breakdown of the hydrogen bonds ia also small, so that
we may use the approximation of Born-Oppenheimer.  Therefore,  it
is possible to consider the operators $B_n, \, B_n^{+}$ assuming
that $\vec q$  and $\vec \omega$ are constant.  At the same we
shall assume that $\gamma $ is small and make all calculations up
to the order $\gamma^2$. In fact, we are considering the
interaction energy given by the term $\gamma \, \vec q_n \cdot R
\vec \sigma_n$, which is smaller than $\epsilon$.  Since $A_n
\propto \gamma$, we shall neglect terms of order $\gamma^2$ in
$(\epsilon^{-1})_{k_1k_2}$ and obtain
$$
 \left( \epsilon^{-1}
 \right)_{k_1 k_2}    = - \displaystyle{
                            \frac{\delta_{k_1 k_2}}
                                 {\displaystyle{ \epsilon - 2 \beta \cos\frac{2 \pi k}{N}}
                                 }
                          }
$$
On setting $q_n = q_n^2  +  i \, q_n^3$ we cast the conformational
energy in the form
$$
    E_{conf} = \displaystyle{ \frac{\gamma}{4}
                  \sum_{n=0}^{N-1} \left( e^{\displaystyle {i \phi_n}} q_n
                                      + e^{\displaystyle{ - i \phi_n}}
                                      q_n^{*}
                                   \right)
               }
        - \displaystyle{ \frac{3 \gamma^2}{16} \sum_{n=0}^{N-1}
                          \displaystyle{ \frac{|D(k)|^2}{\alpha
                             - 2 \beta \displaystyle{ \cos \frac{2 \pi k}{N}}}
                          }
          }
$$
where
$$
 D(k) = \frac{1}{2}  \frac{1}{\sqrt{N}} \sum\limits^{N-1}_{m=0}
     e^{ \displaystyle{ \frac{2 \pi i m k}{N}
       } }
                       \left[
                            \cos \phi_m q^2_m + \sin \phi_m q^3_m
                      \right]
$$
On integrating out the phases $\phi_n$, we find the effective
potential
$$
 U(\vec q ) = E_{conf} =
            \displaystyle{\frac{3 \epsilon^2}{32 \epsilon}
                \sum_{n=0}^{N-1} \, |q_n|^2
            }
$$
or using the integral form
\begin{equation}
 U(\vec q ) = E_{conf} =
    \displaystyle{ \frac{3 \gamma^2}{32 \epsilon d}
                     \int |q_n|^2 \, ds
                 }
 \label{grad}
\end{equation}

Within the framework of the model considered above the excitation
of the hydrogen bonds\label{HWcomp}, i.e their breakdown, is
visualized as an ideal gas, similar to that of magnons.  It is
applied only to regimes far enough from the denaturation, when the
number of the broken hydrogen bonds is small, \cite{Leroy1}.
Nonetheless, it sheds some light on the formation of locally
denaturated regions, or bubbles, of the DNA. It is worthwhile to
recall that G.Manning had suggested that there are two types of
excitations of the DNA; the modes of bending and those of
breathing, \cite{manning}. By now it is generally accepted that
the bending modes are of high frequency, and could be related to
nonlinear phenomena in the DNA. The breathing modes according to
\cite{Leroy1} are in GHz-region, which is at the edge of the
elastic modes of the DNA. If we direct our attention primarily to
bubbles formed by opened base pairs, characteristic time appear to
be by orders of magnitude larger, \cite{Bonnet}. It is also
worthwhile to note that the model of elastic chain of the Pauli
operators, has some bearing on the opening of base pairs and the
breakdown of hydrogen bonds, mentioned on p.\pageref{Hbonds}.

\section{The extended lattice model}
\label{sec:lattice}

As was discussed above, the elastic rod model is not generally
sufficient for describing the conformational dynamics of the DNA.
Important, it does not allow for its intrinsic degrees of freedom
corresponding to the structure of the double helix. To build an
adequate model to the effect, is a  difficult problem, and an
attempt to manufacture it 'from first principles' is doomed to
failure.  Thus, there is a need for drastic simplifications, and
it is necessary  to take into account: (1) the DNA having the two
strands; (2) the base-pairs being linked by the hydrogen bonds;
(3) the helical symmetry of the DNA. The problem still waits its
general solution, but specific cases are nonetheless tractable. In
this section we are going to see what could happen if elastic
modes that can be expected within the rod model, may interact with
internal motions of the double helix.

We  consider  short pieces of the DNA, of several persistence lengths, so that the spatial
conformation of the molecule on the mesoscale , is not of primary importance.  We  focus on the
internal dynamics, trying to accommodate the above requirements through a one-dimensional lattice
model of the DNA. The key point in this respect is the wise choice of dynamical variables that could
give a picture of the DNA dynamics, both simple and adequate. El Hasan and Calladine,
\cite{calladine}, give the framework for such analysis by setting up the scheme for the internal
geometry of the double helix of the DNA. They describe the relative position of one base with respect
to the other in a Watson-Crick base-pair, and also the positions of two base-pairs, by introducing
local frames for the bases and the base-pairs, and translation-slides along their long axes.

We follow the guidelines of paper \cite{calladine}, but aiming at
a qualitative description of the DNA dynamics use a simplified set
of variables. We shall describe the relative position of the bases
of a base-pair by means of the vector $\vec Y$ directed along the
axis of orientation for complimentary bases inside the base pair;
$\vec Y$ being equal to zero when the base-pair is at equilibrium.
The relative position of the base-pairs is described by the
torsional angles $\phi_n$, which give deviations from the standard
equilibrium twist of the double helix. Thus a twist of the DNA
molecule, which does not involve inter-strand motion or mutual
displacements of the bases inside the pairs, is determined by the
torsional angles $\phi_n$ that are the angles of rotation of the
base-pairs about the axis of the double-helix. The twist energy of
the molecule is given by the equation
$$
     \sum_n\, \left[
            \frac{I}{2} \, \dot{\phi}_n^2
            +  \displaystyle{\frac{\tau}{2a^2}} \,
               (\phi_{n+1} - \phi_n)^2
         \right]
$$
in which $I$ is the moment of inertia, and $\tau$ is the twist
coefficient, which for the sake of simplicity and taking into
account the qualitative picture at which we aim, are assumed the
same for all the base-pairs. Inter-strand motions should
correspond to the relative motion of the bases inside the
base-pairs, therefore the kinetic energy due to this degree of
freedom may be cast in the form
$$
  \sum_n\, \frac{M}{2}\, \dot{\vec Y}_n^2
$$
where $M$ is the effective mass of a couple.

For each base-pair we have the reference frame in which z-axis
corresponds to the axis of the double helix, y-axis to the long
axis of the base-pair, x-axis perpendicular to z- and y- axes. At
equilibrium the change in position of adjacent base-pairs is
determined only by the twist angle $\Omega$ of the double helix.
We shall assume $\Omega = 2 \pi / 10$. To determine the energy due
to the inter-strand displacements we need to find the strain
taking into account the constraint imposed by the helical
structure of our system. For this end one may utilize the method
employed by G.Kirchhoff for the twisted rod, that is the covariant
derivative, as was done in paper \cite{kats} for the DNA molecule.
But  a more simple and straightforward approach is possible.

Let us confine ourself only to the torsional degrees of freedom of
the double lattice and assume the vectors $\vec Y_n$ being
parallel to x-y plane, or two-dimensional.
Consider the displacements $\vec Y_n,\, \vec Y_{n+1}$
determined within the frames of the two consecutive base-pairs, n, \, n+1.
Since we must compare the two vectors in the same frame,
we shall rotate  the vector
$\vec Y_{n+1}$ to the frame of the n-th base pair,
$$
  \vec Y^{\, back}_{n+1} =  R^{-1}(\phi)\, \vec Y_{n+1}
$$
Here $R^{-1}(\phi)$ is the inverse matrix
of the rotation of the n-th frame to the (n+1)-one given by the equation
\begin{equation}
  R(\phi) = \left[
          \begin{array}{ll}
                \cos \phi  & - \sin \phi   \\
                \sin \phi  &   \cos \phi
          \end{array}
        \right] \label{rot}
\end{equation}
The matrix $R$ is 2 by 2 since the vectors $\vec Y_n$
are effectively two-dimensional.
Then the  strain  caused by the displacements of the base-pairs
is determined by the difference
$$
   \vec Y^{\, back}_{n+1}  - \vec Y_n
$$
For this argument I am indebted to D.I. Tchertov.

It is important that the angle $\phi$ is given by
the twist angle, $\Omega$, describing the double helix,
in conjunction with  the torsional angles $\phi_n$, so  that
$$
  \phi = \Omega + \phi_{n+1} - \phi_n
$$
Therefore, the energy due to the {\it inter-strand} stress reads
$$
    \sum_n \left\{
           \frac{M}{2}\, \dot{\vec Y_n}^2
           + \displaystyle{\frac{K}{2a^2}} \,
          \left[ R^{-1}(\Omega + \phi_{n+1} - \phi_n)\, \vec Y_{n+1}
            - \vec Y_n
          \right]^2
        \right \}
$$

It corresponds  with the fact that the equilibrium position of the
double helix is the twisted one determined by $\Omega$ and all
$\phi_n$ being equal to zero. We suppose that the size of DNA
molecule is small enough that it can be visualized as a straight
double helix, that is not larger than the persistence length.
Hence the number of base-pairs, $ N \le 150 $, approximately.
Combining the formulas given above we may write down the total
energy of the DNA molecule in the form, \cite{microwave},
\begin{widetext}
\begin{eqnarray}
  {\cal H} &=&
       \sum_n\, \left[
            \frac{I}{2} \, \dot{\phi}_n^2
            +  \displaystyle{\frac{\tau}{2a^2}} \,
               (\phi_{n+1} - \phi_n)^2
         \right] \nonumber \\
       &+& \sum_n \left\{
           \frac{M}{2}\, \dot{\vec Y_n}^2
           + \displaystyle{\frac{K}{2a^2}} \,
          \left[ R^{-1}(\Omega + \phi_{n+1} - \phi_n)\, \vec Y_{n+1}
            - \vec Y_n
          \right]^2
           + \frac{\epsilon}{2}\, \vec Y_n^2
        \right \}    \label{main}
\end{eqnarray}
\end{widetext}
in which  $K$ and $a$ are the torsional elastic constant and the
inter-pairs distance, correspondingly. In summations given above n
is the number of a site corresponding to the n-th base-pair, and $
n= 1,2, \ldots, N $, $N$ being the number of pairs in the segment
of the DNA under consideration. The last term, $\epsilon / 2 \,
\vec Y^2$ accommodates  the energy of the inter-strand {\it
separation} due to the {\it slides of the bases inside the
base-pairs}.

It should be noted that the dynamical  variables $\phi_n$ and
$\vec Y_n$ are of the same order of magnitude, that is the first.
Consequently, preserving only terms up to the third order, we may
transform  Eq.(\ref{main}), so that it takes on the form
\begin{widetext}
\begin{eqnarray}
  {\cal H} &=&
       \sum_n\, \left[
            \frac{I}{2} \, \dot{\phi}_n^2
            +  \displaystyle{\frac{\tau}{2a^2}} \,
               (\phi_{n+1} - \phi_n)^2
         \right] \nonumber \\
       &+& \sum_n \left\{
             \frac{M}{2}\, \dot{\vec Y_n}^2
             + \displaystyle{\frac{K}{2a^2}} \,
             \left[ R^{-1}(\Omega)\, \vec Y_{n+1} - \vec Y_n
             \right]^2
          + \frac{\epsilon}{2}\, \vec Y_n^2
         \right \} \nonumber \\
       &+& \frac{K}{a^2} \sum_n\, (\phi_{n+1} - \phi_n)\,
             \left[ R^{-1}(\Omega)\, \vec Y_{n+1} \times \vec Y_n
             \right]_3       \label{main2}
\end{eqnarray}
\end{widetext}
We have used the fact that the axis of the double-helix is
directed along Oz-axis.

Let us  simplify  Eq.(\ref{main2}) by diagonalizing it with
the help of the unitary transformation
$$
  \vec Y_n = S\, \vec u_n; \qquad
  S = \left[
          \begin{array}{lll}
        \frac{1}{\sqrt{2}}  &   \frac{i}{\sqrt{2}}    \\
        \frac{i}{\sqrt{2}}  &   \frac{1}{\sqrt{2}}
          \end{array}
        \right]
$$
which is  a two by two matrix, for the vectors $\vec Y_n$ and
$\vec u_n$ are effectively two-dimensional, their third coordinates being
equal to zero.
The equation for the energy (\ref{main2}) takes on the form

\begin{widetext}
\begin{eqnarray*}
 \cal{H} &=&
  \sum_n \left[
       \frac{I}{2}\, \dot{\phi}_n^2
       + \frac{\tau}{2 a^2}\, (\phi_{n+1} - \phi_n)^2
     \right] \\
  &+& \sum_n \left[
         \frac{M}{2}\, \dot{\vec u}_n  \cdot \dot{\vec u}^*_n
         + \frac{\epsilon}{2}\, \vec u_n  \cdot \vec u_n^*
                 + \frac{K}{2 a^2}\,
           \left( \mid e^{i \Omega}\, u^1_{n+1} - u^1_n \mid^2
             +   \mid e^{- i \Omega}\, u^2_{n+1} - u^2_n \mid^2
           \right)
          \right] \\
  &-& \frac{K}{a^2} \, \sum_n \, (\phi_{n+1} - \phi_n) \,
    \left[ - i e^{i \Omega}\, u^1_{n+1}\, \stackrel*{u}^1_n
           + i e^{- i \Omega}\, u^2_{n+1}\, \stackrel*{u}^2_n
    \right]
\end{eqnarray*}
\end{widetext}

The star $*$ signifies complex conjugation.

We can further simplify  the equation
for the energy by applying the Fourier transform
given by the equations

\begin{widetext}
\begin{eqnarray*}
  f_n &=& \frac{1}{\sqrt{N}}\, \sum_q\, e^{- i n a q}\, f_q \\
  f_q &=& \frac{1}{\sqrt{N}}\, \sum_{n = -N/2}^{n = + N/2}\,
                  e^{ i n a q}\, f_n \quad
       q = \frac{2 \pi}{N a}\, m; \quad
       m = 0, \pm 1, \pm 2, \ldots, \pm \frac{N}{2};
\end{eqnarray*}
\end{widetext}

It is important that after the Fourier transform the variables $\vec u_n$
verify the following equations for their complex conjugates
\begin{equation}
  \stackrel*{u}^1_q  = i u^2_{-q}, \quad
  \stackrel*{u}^2_q  = i u^1_{-q}            \label{cconj}
\end{equation}
The equation for the energy can be cast in the form

\begin{widetext}
\begin{eqnarray}
 \cal{H} &=&
  \sum_q \left[
       \frac{I}{2}\, \dot{\phi}_q\, \dot{\phi}_q^*
       + \frac{\tau}{2 a^2}\, \sin^2\, \frac{a q}{2}\, \phi_q\, \phi_q^*
     \right]   \nonumber \\
  &+& \sum_q \left[
         \frac{M}{2}\, \dot{\vec u}_q  \cdot \dot{\vec u}^*_q
         + \frac{\epsilon}{2}\, \vec u_q  \cdot \vec u_q^*
                 + \frac{2 K}{a^2}\,
           \left(
             \sin^2\, \frac{\Omega - a q}{2} \,
                   u^1_q \stackrel*{u}^1_q
              +  \sin^2\, \frac{\Omega + a q}{2} \,
                   u^2_q \stackrel*{u}^2_q
           \right)
          \right]    \nonumber \\
  &+& \frac{K}{a^2} \, \sum_{q' q'' }\, i \frac{e^{- i aq}}{\sqrt{N}}\,
                    \phi_{q'}\,
    \left[ - e^{i \Omega}\, u^1_{q''}\, \stackrel*{u}^1_{q'+q''}
           + e^{- i \Omega}\, u^2_{q'}\, \stackrel*{u}^2_{q'+q''}
    \right]       \label{main3}
\end{eqnarray}
\end{widetext}

\noindent
in which
$$  q_* = \Omega / a$$

The above equation serves as well a Hamiltonian that describes the dynamics of a molecule of the DNA,
within the framework of the present model, up to terms of the third order. It is worth noting that the
latter is based on the assumptions given above concerning the basic structural properties of the DNA,
that is it takes into account its two-strand structure, the lattice formed by the base pairs, and the
helical symmetry. The specific feature of the Hamiltonian is the presence of the interaction term that
describes three-wave interaction, \cite{microwave}, and may result in resonance. We shall utilize the
fact  for deriving the parametric maintenance of the $u_q$ modes, i.e. the HBS (hydrogen-bond-stretch)
modes, (see below).

One can obtain, in the usual way, the equations of motion for
$u_q^{\alpha}, \, \alpha = 1,2$ and $\phi_q$, from the equation
for the energy indicated above. The essential point is the effects
of dissipation, which are due to ions  in the close neighborhood
of the molecule and water effects, see \cite{zandt}. The
dissipation could be accommodated by writing down terms linear in
$\dot{u}_q^{\alpha},\,  \dot{\phi}_q$. We shall take into account
external force, or torque ${\cal T}_q$, only in the equation for
$\phi_q$, for it corresponds to external degrees of freedom of our
model. Thus, the equations of motion can be cast in the form

\begin{widetext}
\begin{eqnarray}
  \ddot{u}^{\alpha}_q + \omega^2_{\alpha\,q} u^{\alpha}_q
  + \gamma_u\, \dot{u}^{\alpha}_q
  &+&
  \frac{4 K \sin \Omega }{Ma^2\, \sqrt{N}}
     \sum_{q'}\, e^{- i a q'}\, \phi_{q'} u^{\alpha}_{q-q'} =  0,
         \label{motion_f1} \\
  \ddot{\phi}_q + \omega^2_q\, \phi_q
    + \gamma_{\phi} \dot{\phi}_q
  &+&
    i \frac{4 K \sin \Omega \, e^{ i a q}}{I a^2\, \sqrt{N}}
     \sum_{q'}\,  u^1_{q'} u^2_{q-q'}
     = {\cal T}_q   \label{motion_f}
\end{eqnarray}
\end{widetext}

\noindent
Here

\begin{widetext}
\begin{equation}
  \omega^2_{\alpha q} = \frac{4 K}{M a^2}\,
     \sin^2 \frac{\Omega + (-1)^{\alpha} aq}{2}\, +\, \frac{\epsilon}{M},
  \quad
  \omega^2_q = \frac{4 \tau}{I a^2}\, \sin^2\frac{aq}{2}
  \label{freqs}
\end{equation}
\end{widetext}

\noindent are the dispersion laws for the fields $u_q^{\alpha},\,
\alpha=1,2$, and $\phi_q$. We see that the spectrum of $\phi_q$
has a typical acoustic character, whereas that for $u^{\alpha}_q$
has a local minimum determined by the helical twist, $\Omega$.
Thus, the spectrum of our model is in qualitative agreement with
conclusions of \cite{proh1}.  The specific nature of the torque is
to be specified elsewhere, for the moment, we shall consider
general dynamical phenomena to which the torque may be conducive.

Let us suppose that for one thing the amplitudes of the HBS-modes
given by $u^{\alpha}_q$ be so small that the quadratic term in
Eq.(\ref{motion_f}) can be neglected, and for another the
external torque ${\cal T}_q$ be appreciable enough to maintain the
vibration of the torsional mode $\phi_q$.  Thus, we can visualize
the latter as a pump mode that interacts with the HBS-mode
$u^{\alpha}_q$ through the non-linearity in
Eq.(\ref{motion_f1}). We shall confine ourself to the case of
the torque ${\cal T}_q$ being non-zero only at $q = q_*$ and
having a frequency $2\omega$. Therefore, the forced wave, or the
pump wave for the HBS-mode, has the form
\begin{equation}
  \phi_{q^*} = e^{i 2 \omega t}\, \Phi\, \delta_{qq^*}, \quad
  \phi_{-q^*} = e^{- i 2 \omega t}\, \Phi^*\, \delta_{-qq^*}
  \label{ppp}
\end{equation}
To obtain larger values for the pump wave, $\phi_q$, the resonance
condition
$$
   \omega_{q^*} = 2 \omega
$$
should be verified, even though the resonance behavior of the
torsional  $\phi_q$-mode itself could be attenuated by
dissipation, i.e. it may be a mode of small amplitude.

The equations of motion for $u^{\alpha}_q$ in the pumping regime read
\begin{widetext}
$$
  \ddot{u}^{\alpha}_q  + \omega^2_{\alpha q} u^{\alpha}_q
               + \gamma_u \dot{u}^{\alpha}_q
  + \frac{2K}{Ma^2}\frac{\sin \Omega}{\sqrt{N}}
      \left(  A\, e^{i 2 \omega t}\, u^{\alpha}_{q - q*}
        + A^*\, e^{- i 2 \omega t}\, u^{\alpha}_{q + q*}
      \right)
  = 0
$$
\end{widetext}
here
$$
   A = e^{-i \Omega}\, \Phi
$$
Note that the momentum conservation in the q-values is preserved,
as required by the three-wave interaction.
The equations indicated above can be cast in the matrix form
\begin{equation}
   \ddot{\vec u}_{\alpha} + \hat{\omega}_{\alpha}^2 \vec u_{\alpha}
              + \gamma_u \dot{\vec u}_{\alpha}  =
    \left( e^{i 2 \omega t}\, {\cal K} +
       e^{- i 2 \omega t}\, {\cal K}^{+}   \right) \vec u_{\alpha}
    \label{matr}
\end{equation}
where ${\cal K}$ and ${\cal K}^{+}$ are hermitian conjugate, and
$$
  {\cal K}^{+}{\cal K} = {\cal I}\, \left( \frac{2 K}{M a^2}
                   \frac{\sin \Omega}{\sqrt{N}}
                 \right)^2\, \mid A \mid^2, \qquad
                 {\cal I}_{ij} = \delta_{ij}
$$
It is worth noting that  Eq.(\ref{matr}) is a kind of matrix
Mathieu equation. In fact, we can apply to it Rayleigh's  method
for studying parametric resonance, \cite{Rayleigh}. For this end
let us look for the solution to  Eq.(\ref{matr})  in the form
of a series
$$
  \vec u(t)  =
  \vec A_1\, e^{i \omega t} + \vec B_1\, e^{- i \omega t} +
  \vec A_3\, e^{i 3 \omega t} + \vec B_3\, e^{- i 3 \omega t} + \ldots
$$
On substituting the expression given above into
Eq.(\ref{matr}) and preserving only the terms corresponding
to $e^{\pm i \omega t}$, we obtain the equations
\begin{eqnarray*}
  \left[ (- \omega^2 + i \gamma_u \omega)\, {\cal I} + \hat{\omega}_{\alpha}^2
  \right]\, \vec A_1 &+& {\cal K}\, \vec B_1 = 0  \\
  \left[ (- \omega^2 - i \gamma_u \omega)\, {\cal I} + \hat{\omega}_{\alpha}^2
  \right]\, \vec B_1 &+& {\cal K}^{+}\, \vec A_1 = 0
\end{eqnarray*}

The compatibility condition of the equations indicated above can be cast
in the form of determinant for the block matrix
\begin{equation}
  det\, \left[
   \begin{array}{cc}
     \hat{\omega}_{\alpha}^2 - \omega^2 + i \gamma_u\, \omega
      &  {\cal K} \\
      {\cal K}^{+}
      & \hat{\omega}_{\alpha}^2 - \omega^2 - i \gamma_u\, \omega
   \end{array}
   \right] = 0
   \label{rc}
\end{equation}
Here $\hat{\omega}^2$ is the matrix of frequencies given by
Eq.(\ref{freqs}), and $\omega^2$  and $\gamma_u\, \omega$ are
the scalar ones. We can transform  Eq.(\ref{rc}) into a more
amenable form. Let us notice that it is equivalent to the equation

\begin{widetext}
$$
det \,  \left \{
\left[
   \begin{array}{cc}
     \hat{\omega}_{\alpha}^2 - \omega^2 + i \gamma_u \omega
      &  {\cal K} \\
      {\cal K}^{+}
      & \hat{\omega}_{\alpha}^2 - \omega^2 - i \gamma_u \omega
   \end{array}
   \right]
\left[
   \begin{array}{cc}
    {\cal I}
      & - (\hat{\omega}_{\alpha}^2 - \omega^2 + i \gamma_u  \omega)^{-1}    \\
    0
      & \rho^2  {\cal K}^{+}
   \end{array}
   \right]
\right \}
   = 0
$$
\end{widetext}

\noindent
in which
\begin{equation}
   \rho =  \frac{Ma^2}{2K} \frac{\sqrt{N}}{\sin \Omega} \label{rho}
\end{equation}
and the matrices ${\cal K}^{+}$ and ${\cal K}$ verify the equation
$$
  - {\cal I} + \rho^2 \, {\cal K} \, {\cal K}^{+} = 0
$$
We have used the fact that for the range of frequencies we are considering,
the matrix
$$
  \hat{\omega}_{\alpha}^2 - \omega^2 + i \gamma_u \, \omega
$$
is not degenerate. Therefore, the equation given above is equivalent to
the following one

\begin{widetext}
$$
 det \left[
       - {\cal I} + \rho^2 \,
             ( \hat{\omega}_{\alpha}^2 - \omega^2 + i \gamma_u \, \omega )
               {\cal J}
             ( \hat{\omega}_{\alpha}^2 - \omega^2 - i \gamma_u \, \omega )
               {\cal J}^{+}
     \right] = 0
$$
\end{widetext}

\noindent
in which the matrix ${\cal J}$ is given by
$$
  {\cal J}_{qq'} = \delta_{q'\, q - q_*}
$$
We may cast the last equation into the form

\begin{widetext}
\begin{equation}
 (\omega_{\alpha q}^2 - \omega^2 - i \gamma_u \, \omega )
 (\omega_{\alpha\, q-q_*}^2 - \omega^2 + i \gamma_u \, \omega )
 - \left( \frac{2 K}{M a^2} \frac{\sin \Omega}{\sqrt{N}}
   \right)^2 \, |A|^2  =0
   \label{rc2}
\end{equation}
\end{widetext}

\noindent which is quite similar to the usual condition for
parametric resonance. Solutions to  Eq.(\ref{rc2}) are
generally complex and therefore correspond to attenuated regimes.
But there is a specific wave number, $q_{res}$, for which the
solution gives the real frequency $\omega$, and it is easy to see
that it should satisfy the constraint
\begin{equation}
    \omega^2_{\alpha q-q_*} =  \omega^2_{\alpha q}, \qquad \mbox{at }\qquad q = q_{res}
    \label{resQ}
\end{equation}
Thus, we may cast the condition for parametric resonance
in the familiar form, \cite{Rayleigh},
\begin{equation}
  (\omega^2 - \omega^2_{\alpha q_{res}})^2
  + \gamma^2\, \omega^2
  - \left(  \frac{2 K}{M a^2} \frac{\sin \Omega}{\sqrt{N}}
    \right)^2\, |A|^2  = 0
   \label{pr}
\end{equation}

The existence of the parametric resonance discussed above presupposes that  the bands of the torsional
acoustic (TA), the $\phi_q$ modes of the present paper, and the hydrogen-bond-stretch (HBS) modes,
that is the $u_q^{\alpha}$ modes,  of the DNA interpenetrate each other. So far, the knowledge of the
spectra of the DNA vibrations relies mainly on the computer analysis within the framework of quantum
chemistry, \cite{proh1}. According to these results, it is accepted that the conformational dynamics
of the DNA is confined  to elastic vibrations of the DNA molecule in the range of $10^9 \pm 10^{12} \,
Hz$, \cite{star}. Kim and Prohofsky, \cite{proh1}, claim that the region comprises two domains, which
correspond with different degrees of freedom of the molecule: (1) acoustic modes, which do not involve
the hydrogen bonds; (2) modes that stretch the hydrogen bonds between the base-pairs (the HBS modes).
Local minimum of the frequency is characteristic of the HBS-modes, \cite{proh1}; its position
depending on the choice of the band. The vibrations of the DNA, which are ascribed to the inter-strand
modes, were observed in the low-frequency Raman scattering, \cite{urabe}, \cite{urabe2}, and the
Fourier-transform infra-red absorption experiments, \cite{powell}. Globus et al, \cite{globus1},
report the existence of internal modes generated by the interaction of artificial DNA-type molecules
with electromagnetic radiation in sub-millimetre range. It should be noted that the type of modes
observed depends on the kind of DNA samples, i.e. in aqueous solutions, or films and filaments,
\cite{globus1}.  The experimental data, \cite{star}, is not conclusive as to the relative positions of
the acoustic and the HBS modes. The inter-strand and the acoustic modes of the DNA are  alleged to be
overdamped, \cite{proh1}. But the opinion of scientific community in this respect is not unanimous.
First of all, they are observed and measured, fairly well, in experiment, \cite{edwards},
\cite{urabe}, \cite{urabe2}, \cite{edwards}.  Second, according to paper \cite{georghiou} it is the
modes related to motion of the sugar-phosphate backbone that are overdamped, that is the $\phi_q$
modes discussed above, whereas the inter-strand modes, or the above $u_q^{\alpha}$ are not. Third,
Davis  and VanZandt, \cite{zandt}, had shown that the dissipative effects suffered by the modes are
greatly diminished in case the GHz-frequency range is taken into account; in the region it is
necessary to employ the so-called Maxwell model of hydrodynamics, or the Leontovich theory,
\cite{fabelinsky}. Thus, we see that there are serious arguments against the inter-strand modes being
overdamped and absent.

We are in a position to assess the action of mw-radiation on the molecule of the DNA. The question
which has aroused controversy during the last fifteen years. We wish to make it clear that there is no
"scare-mongering", but merely {\it a suggestion to employ microwave radiation for studying the
biophysics of the DNA}. In fact, we feel that the technic of Raman scattering, which had been
successfully used for detecting the inter-strand modes, in conjunction with the microwave radiation
could be instrumental in studying intrinsic motions of the DNA.

The key point of the theoretical analysis of the interaction of
electromagnetic radiation with the DNA is accommodating the fact
that the wavelength of radiation is by many orders of magnitudes
larger than the characteristic size of the region of the molecule
involved in the process. It was Chun-Ting Zhang, \cite{zhang}, who
suggested a mechanism to overcome this difficulty. The main point
of Zhang's argument is that the helical configuration of the
electric dipoles corresponding with the base-pairs makes the
interaction of the dipole $\vec P$  and the field $\vec E$
$$
  U = -\vec P \cdot \vec E
$$
dependent on angle.
Therefore, different  torsional momenta are applied at the  base-pairs.
The equation for the energy of interaction between the dipoles of DNA
and an incident micro-wave reads
$$
 -  \sum_n\, \vec E \cdot R(n \Omega + \phi_n) \vec P_{o}
$$
Here $R(n \Omega + \phi_n)$ is the rotation matrix given by Eq.(\ref{rot}), and $\vec P_o$ is the
dipole at site $n=0$. Consequently, even though on the molecular scale the radiation has a plane wave
configuration, it still twists the DNA molecule about the axis of the double-helix. Since the momenta
changes periodically in time with the incident wave, the irradiation results in a periodic stress that
may produce elastic vibrations in the DNA molecule.  Zhang suggested that the force may generate
resonance vibrations,  resulting in  a cross-over mechanism which takes up initial torsion excitations
and transforms them into longitudinal acoustic vibrations.

In the present paper we will try to combine Zhang's mechanism,
\cite{zhang}, and the excitations of the double-helix studied by
Prohofsky and Kim, \cite{proh1}, with the view of generating
inter-strand waves in the DNA  by mw-irradiation. In contrast to
the original idea by Zhang, we do not utilize a cross-over into
longitudinal acoustic vibrations, but employ the interaction
between torsional oscillations and the inter-strand ones, i.e. the
three-wave, given by  Eq.(\ref{main3}).

The main point is that
by expanding the rotation matrix $R(n \Omega + \phi_n)$ in the angles
$\phi_n$ and keeping only the first order terms, we may cast
Zhang's interaction in the form
\begin{equation}
  {\cal H}_Z = - \sum_n \, \phi_n \, (\vec E \times \vec P_n )_3
  \, + \, const, \quad \vec P_n = R(n \Omega) \, \vec P_o
  \label{Zinter}
\end{equation}
in which $\vec P_o$ is the dipole vector at site $n = 0$. Next, by
using  Eq.(\ref{rot}) for the matrix $R(n \Omega)$ and
neglecting the constant term we may cast  Eq.(\ref{Zinter})
in the form

\begin{widetext}
$$
 {\cal H}_Z = \frac{1}{2}
          \sum_n \, \phi_n \,  \left \{
         e^{i n \Omega} \, [   (\vec E \times \vec P_o)_3
                    - i (\vec E \cdot \vec P_o) ] \,
        + \, e^{- i n \Omega} \, [   (\vec E \times \vec P_o)_3
                      + i (\vec E \cdot \vec P_o) ]
          \right \}
$$
\end{widetext}

On applying the Fourier transform for the $\phi_n$, and utilizing
the equation
$$
  \frac{1}{N} \, \sum_n \, e^{i(\Omega \pm aq) n} =
     \delta_{\Omega, \pm aq}
$$
we obtain the following expression for Zhang's interaction

\begin{widetext}
$$
 {\cal H}_Z = \frac{N}{2} \left \{
         \phi_{q = \frac{\Omega}{a}}  \, [ (\vec E \times \vec P_o)_3
                      - i (\vec E \cdot \vec P_o) ] \,
        + \, \phi_{q = - \frac{\Omega}{a}} \, [ (\vec E \times \vec P_o)_3
                      + i (\vec E \cdot \vec P_o) ]
          \right \}
$$
\end{widetext}

Hence, the torque ${\cal T}_q$ in  Eq.(\ref{motion_f})
corresponding to ${\cal H_Z}$ is given by the equation
\begin{equation}
  {\cal T} = \frac{{\cal Z}}{I}   \, \delta_{q, - q^*} + \,
         \frac{{\cal Z^*}}{I} \, \delta_{q,  q^*} \qquad
       q_* = \Omega / a
       \label{torqueZ}
\end{equation}
in which
$$
  {\cal Z} =  \frac{N}{2}\,
          \left[
         ( \vec E \times \vec P_o)_3 + i (\vec E \cdot \vec P_o)
          \right]
$$
It should be noted that $\pm q_*$ are the local minima of the
HBS-modes. From  Eq.(\ref{freqs}) we infer that $q$ verifies
the constraint given by  Eq.(\ref{resQ})  reads
\begin{equation}
  q_{res} = \frac{3}{2}\, q_{*}   \label{pump}
\end{equation}
It is worth noting that the wave numbers $q_{*}$ and $q_{res}$
correspond to the wavelengths of one and $\frac{2}{3}$ turns of
the double-helix.

Equations given above provide an opportunity for making numerical,
order of magnitude, estimates, which enable us to assess the
effect of mw-radiation on the HBS-modes.  From
Eq.(\ref{torqueZ}) we infer that the torque ${\cal T}$ has
the size
$$
  {\cal T} \propto e^{2 i \omega t}\, E \, P
$$
where $E$ and $P$ are the external field and the dipole moment of
the base-pair, respectfully. Next, suppose that the resonance condition
$$
   \omega_q  = 2 \omega, \qquad q = q_* = \frac{\Omega}{a}
$$
be true, so that the action of the radiation on the torsional
modes should be the largest possible.  Then the amplitude of the
pumping wave, $\phi_{q_*}$, according to
Eq.(\ref{motion_f}), is of the order
\begin{equation}
   \Phi \propto \frac{\sqrt{N}}{I}\, \frac{E P}{2 \omega \gamma_{\phi}}
   \label{Phi}
\end{equation}
Next, we turn to Rayleigh's condition for the parametric resonance
of the HBS-mode given by  Eq.(\ref{pr}). For the pumping wave
corresponding to  Eq.(\ref{Phi}), it gives
$$
  ( \omega^2 \, - \, \omega^2_{\alpha q_*})^2 +
  \gamma^2_u \, \omega^2 \approx
  4 \left ( \frac{K \sin \Omega}{M a^2}\,
        \frac{E P}{I \gamma_{\phi}}  \right )^2
$$
Hence we have the threshold

\begin{equation}
   \gamma_u \, \gamma_{\phi} \le
   \frac{2 K \sin \Omega}{M a^2 \omega^2} \, \frac{E P}{I}
   \label{effect}
\end{equation}

\noindent which is {\em the condition that the energy supplied to
a DNA molecule is greater than that dissipated}, so that the
maintaining of the HBS-mode can take place. We suppose that the
frequency of the HBS-modes, as given by  Eq.(\ref{freqs}), is
generally determined by the gap term $\epsilon / M$ in the
equation for $u_{\alpha q}$ and the first factor in
Eq.(\ref{effect}) does not differ much from unity. It
signifies that the energies of the inter-strand separation per
base-pair and the twist of the relative positions of the two
adjacent base-pairs, should be comparable. At any rate, the
hypothesis appears not to contradict the data reproduced in paper
\cite{proh1}. If so, we could have the estimate for the
dissipative constants, at least by orders of magnitude,
\begin{equation}
   \gamma_u \, \gamma_{\phi} \le  \frac{E P}{I}
   \label{estimateE}
\end{equation}
On utilizing the relation
$$
   E \propto 2 \, \sqrt{\frac{\pi \, S}{c}}
$$
which follows from the expression for Pointing's vector
$$
   \vec S = \frac{c}{4 \pi} \vec E \times \vec H,
$$
in which $c$ is the velocity of light, we cast the estimate given
by  Eq.(\ref{estimateE}) in the form
\begin{equation}
   \gamma_u \, \gamma_{\phi} \le 2 \, \frac{P}{I} \,
                 \sqrt{\frac{\pi\, S}{c}}
   \label{estimateW}
\end{equation}
in which $S$ is the power density of the interaction.
If we assume
$$
P \propto 1 \, Debye \quad \mbox{or} \quad  10^{-18} \, CGS
$$
and the inertia coefficient $I \propto 10^{-36} \, gr \, cm^2$,
corresponding to the mass of the base-pair $\propto 10^{-22} \,
gr$, and the size $ \propto 10 \, \AA $,  then  for the power
density $S \propto 100 \, mW/cm^2$, we have
$$
   \gamma_u \, \gamma_{\phi} \le 10^{16} \, Hz^2
   \quad \mbox{or} \quad
   \gamma_u , \gamma_{\phi} \le 10^8 \, Hz
$$
The estimate suggests that the effect produced by mw-radiation is
to be looked for at the edge of the GHz zone, for in this case the
requirement on the line-width is less stringent. It should be
noted that the crucial point in assessing the feasibility of
experiments on mw-irradiation of the DNA, and its possible
influence, is the part played by ambient solvent and ions
contained in it. In fact, the irradiation may result in just
heating the solvent, so that the dissipation due to the ions takes
up all effects on the molecules of DNA. Generally, the thin
boundary layer of water and ions close to the DNA-molecule may
have an important bearing on the dynamics initiated by the
incident mw-radiation and result in the overdamping of the
molecule's torsional oscillations.

Davis and VanZandt, \cite{zandt}, put forward arguments that the
ions contained in a layer close to the DNA molecule should have an
influence small enough to allow the survival of the effect due to
mw-irradiation. The part played by the dissipation caused by water
is more subtle.

The current arguments,\cite{adair}, about the overdamping of the DNA elastic modes, rely on the Stokes
law for  frictional force, $F= 6 \pi \eta R v$, for a sphere of radius $R$ moving in a fluid of
viscosity $\eta$ at speed $v$. It is important that for the specific case of the DNA it should involve
the GHz region of frequencies. But,   the classical hydrodynamics, that is the Navier-Stokes theory,
breaks down in the region, as can be inferred from the phenomenon of light-scattering in liquids,
which is characterized by the triplet structure: the central Rayleigh line, $\nu$, due to the elastic
scattering, and the Mandelstam-Brillouin doublet, $\nu \pm f$, of the inelastic one; $f$ being the
frequency of elastic waves in liquid. The classical hydrodynamics gives the width of a line in the
Mandelstam-Brillouin doublet larger than the distance between this line and the maximum of the central
line of the triplet, so that the discrete triplet  structure should not be observable; in fact, it is,
\cite{fabelinsky}. Mandelstam and Leontovich, \cite{fabelinsky}, brought about the solution to this
problem by using the relaxational theory of hydrodynamics in which liquid is considered as a viscous
elastic medium characterized by a coefficient $\eta$ of viscosity and a shear modulus $G$, the
so-called Maxwell model.  In fact, the theory also takes into account effects of anisotropy,
\cite{fabelinsky}.  It predicts that in the region of hypersound, a few GHz or more,  the attenuation
coefficient for sound waves , $\alpha_{\eta}$, ceases to depend on frequency $\omega$, whereas in the
low frequency classical region, in which the Stokes law is valid, the dependence reads $\alpha_{\eta}
\sim \omega^2$. Davis and VanZandt, \cite{zandt}, used the approach of the Maxwell theory to find
estimates for the damping of the DNA elastic modes, taking into account the first and the second
hydration layers, and the quasi-crystallin structure of water in a neighborhood of the DNA. They found
the attenuation to be two orders of magnitude smaller than that given by the Stokes law.

In fact, there is a further reason for rejecting the approach
based on the Stokes law. The water molecules form hydration shells
of DNA, \cite{tao}, \cite{tao2}. The primary hydration shell
comprises the water molecules immediately adjacent to the DNA,
about 20 molecules per nucleotide pair, which constitute a medium
different from bulk water. The secondary hydration shell is
generally considered to be similar to bulk water. But, at the
spatial scale of the diameter of the DNA-molecule, that is several
tens $ \AA $, the water in the second hydration shell, is hardly a
condensed medium. Indeed, in this case  one should have
accommodated its local quasi-crystalline structure, described by
the icosahedral model, \cite{chaplin}, \cite{mull}, which is to
result in sophisticated dynamical equations. The conclusion is
that, presently, it is difficult, if possible at all, to construct
accurate theoretical estimates for the attenuation of DNA-modes.

From the experimental point of view the situation is more
advanced.  The DNA helical modes were observed in the experiments
on the Raman,\cite{urabe}, \cite{urabe2}, \cite{tao}, and the
far-infrared, \cite{powell}, scattering.  Therefore, one may
suggest that the attenuation effects due to viscosity should not
preclude  elastic modes of the DNA. At the same time small relaxation
times for damping between the DNA and the first hydration layer,
of order of several tens ps, (see \cite{tao2}) should result in
the double helix of DNA concerted motion with surrounding layer of
water. The circumstance could be accommodated within the framework
of the semi-phenomenological model of the present paper. In fact,
the DNA molecule and its first hydration layer still form a helix
structure, and the mutual motion of constituent bases of a pair
together with hydration water molecules could be described with
the field $\vec Y$. Of course, the values of the model's
constants, $K, \tau$, should be changed, and for the time being
there is lack of information as to their size.

It is also worth noting that the effects of dissipation in aqueous
solutions, where a certain form of the Stokes law could be
possible, and in films, or fibers, should be quite different. So
far there has been no comprehensive theoretical analysis of the
dissipation, which would allow for comparing the DNA dynamics in
solutions and in films . Nonetheless, the interplay of internal
vibration modes  and sub-millimeter electromagnetic irradiation
was registered in paper \cite{globus1}, using Fourier transform
spectroscopy and films of the double-stranded homopolymers
poly[A]-poly[U] and poly[C]-poly[G] . Employing the concept of
normal modes, or oscillators, of macromolecules, developed earlier
for proteins, \cite{go}, and used later for DNA, \cite{zakrz},
Globus et al, \cite{globus1}, made a numerical simulation of their
experimental results, and thus obtained an estimate for the
relaxational parameter $\gamma$, which has the meaning of
oscillators dissipation. It turned out that in the range of
frequencies several $10 cm^{-1}$, the best fit for $\gamma$ is
less than $1 cm^{-1}$, depending on the conformation of an
external electric field and a sample. This value of $\gamma$ is
too large for  Eq.(\ref{estimateW}), but the region of
frequencies studied in \cite{globus1} is far from the edge of GHz
region, so that one may consider the question of acceptable rate
of dissipation as still open, and suggest that studying the
effects of mw-radiation on the DNA modes may be instrumental for
understanding the phenomenon.

We see that the elastic dynamics of the double helix could have
enough structure for providing a means  for stretching the
hydrogen bonds of the base-pairs of DNA, or generating the
HBS-modes. If the vibrational modes of the DNA are not overdamped
by the ambient solvent, and the balance between  energies supplied
and dissipated is favourable, the maintenance of the HBS-modes
could be expected at the edge of the HBS-zone. The best technique
for studying the H-bond stretching still remains the Raman
spectroscopy on which certain improvements have been made (see
\cite{moliveanu} and references therein). Thus, the HBS-modes, and
also the breathing modes, are well accessible from the
experimental point of view.

The choice of specific means for generating torsional excitations
of the DNA is important and interesting. In this paper we have
envisaged mw-irradiation of the DNA. In case the interpenetration
of the acoustic and the HBS-modes takes place, mw-radiation could
maintain the HBS-modes, if the power density is sufficiently
large, $100 \, mW/cm^2$ or more. It is important that there is no
need for long exposures of the sample to the radiation. If the
effect be sufficiently pronounced, it may result in the formation
of the bubbles of broken H-bonds. At this point it is worth noting
that our estimate for the critical power density, $100 \,
mW/cm^2$,  is by orders of magnitude larger than that officially
prescribed, i.e. $0.2 \, - \, 0.1 \, mW/cm^2 $.

\section{Proton tunneling inside the hydrogen bonds of the DNA}
\label{sec:tautomer}

In the previous sections we have just considered a few specific situations which, nonetheless,
indicate that the use of simple models and rough approximations is not sufficient for the study of
conformations accessible for the DNA.  The main conclusion is that a molecule  of the DNA has an
intrinsic structure that should be accounted for. Similar situations happen in continuum mechanics
when it is necessary to consider a medium having an internal structure.  But, the problem of the DNA
is more sophisticated since the system under investigation is not a continuum medium, in fact, it is a
macromolecule. Looking at the problem squarely, we  have to acquiesce that we should work within the
framework of nonlinear elasticity theory, if we wish to follow in this way. But the situation is still
even more difficult owing to the necessity to allow for the intrinsic structure of 'the material'.
Thus, at first sight the problem does not look tractable.

It is reasonable to diminish the scope of regimes under the
investigation and confine ourself mainly to problems that could be
treated within the framework of the internal, or inter-strand,
dynamics while assuming that deformations of the molecule of DNA
are small on the mesoscale, that is a few persistence length.  The
assumption does not exclude the presence of external influences,
as can be inferred from results of Section \ref{sec:elrod}. There
is another point we have to take into account: the dynamics of
proton inside the hydrogen bonds between the base pairs. In
contrast to the inter-strand modes of the DNA, they are
essentially quantum modes. Therefore, we need a model that
combines classical elasticity of the double helix and quantum
dynamics of the protons.

The model of this kind had been worked out by Davydov,
\cite{Davy}, for the needs of protein dynamics.  It has been used
for the dynamics of protons in the DNA in papers \cite{gv},
\cite{kats}.  It is important that the Davydov model has enough
structure to accommodate both  the inter-strand modes and the
proton tunneling. The interaction between the two dynamics is a
subject of great controversy, and whether it really takes place,
or is merely speculative, depends  on values of the elastic
constants of the DNA, which are by no means precisely known. But
if the numerical values are favourable, we may expect an
interesting interplay between these forces. In what follows, we
shall try to see what consequences could be inferred in this case.

By changing both positions and mutual orientations of the base
pairs, the inter-strand dynamics of the DNA should deform the
hydrogen bonds  between them, and thus have some bearing upon the
protons. It should be noted that a proton effecting a hydrogen
bond between two bases of the DNA, do not have a unique position
of equilibrium. Under ordinary circumstances it occupies a
position that corresponds to the bases being in the {\it amino},
or {\it keto}, forms for  {\it adenine-guanine}, and {\it
cytosine-thymine}, respectfully.  The change in position of the
proton result in the transition $amino/ket \rightarrow imino/enol$
of the DNA base pairs.  Does the transition influence  elastic
properties of the molecule of the DNA ?  At this point we again
come across the interplay between the microscopical and
macroscopical dynamics of the DNA, and its significance for the
proton transport.  The latter is of primary importance for the
DNA, for among other things it could be  a cause of spontaneous
mutations. It also raises a question of whether irradiation with
electromagnetic waves could result in generating the inter-strand
modes, deforming  the dynamics of proton tunneling, and causing
genetic effects.

\begin{figure}
  \begin{center}
     \includegraphics*[width=13cm, height=14.219cm]{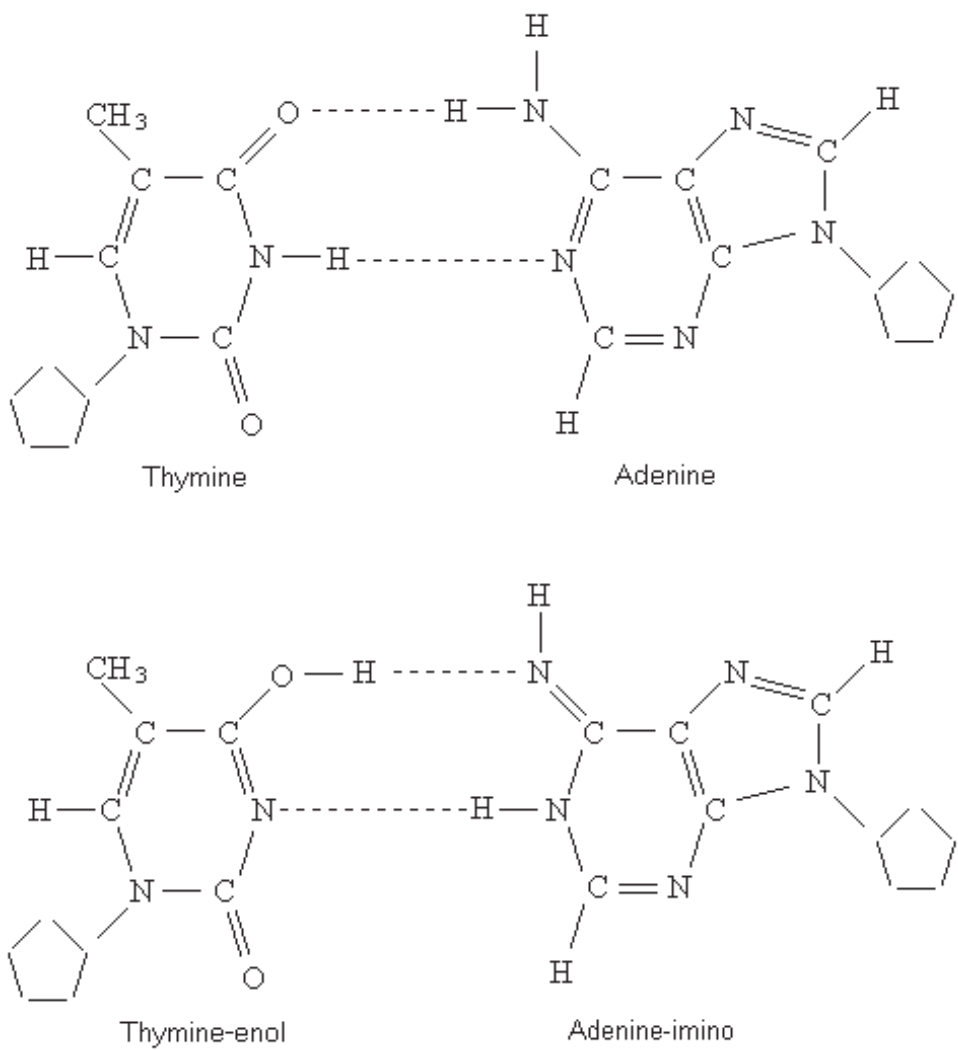}
     \caption{Pairing of Thymine-Adenine in the normal
     keto/amino forms in comparison with the tautomer-shifted enol
     and imino forms.
     \label{fig1}}
  \end{center}
 \end{figure}

Recall that according to the Watson-Crick hypothesis\cite{CW1},
the double helix of the DNA molecule comprises the two strands
linked together by purine-pyrimidine base-pairs of adenine-thymine
(AT) and guanine-cytosine (GC), the four chemicals A,T,G,C
existing in various isomeric forms, or tautomers, that may change
into one another (see FIG.~\ref{fig1}, for example).
Under ordinary conditions the equilibrium shifts towards the
amino-form for adenine and guanine, and the keto-form for thymine
and cytosine. But the imino-form for the adenine and cytosine, and
the enol-form for guanine and thymine are also possible, even
though rare; in fact, they correspond to concentrations of
$10^{-4}$  to $10^{-5}$ moles/liter.\cite{Saenger} The
implications wrought by the tautomeric transitions are important
in that the sequence of base-pairs constitutes the genetic
information of cell, so that exact copies of the DNA should be
produced during the replication. In fact, the complimentarity
between the bases may change if a tautomeric transition takes
place, and other combinations become possible,
\begin{eqnarray}
 A_{imino} \leftarrow  \rightarrow  C & , & \qquad
 A \leftarrow  \rightarrow  C_{imino}   \\  \label{corr1}
 G_{enol} \leftarrow  \rightarrow   T  & , & \qquad
 G \leftarrow  \rightarrow   T_{enol} \nonumber
\end{eqnarray}
in contrast to the usual and stable ones
$$
      A \leftarrow  \rightarrow   T \qquad , \qquad
      G \leftarrow  \rightarrow   C
$$
An opportunity for generating  "unnatural"  pairs arises also from
the tunneling of protons in hydrogen bonds (see FIG.~\ref{fig1}),
which results in the formation of the pairs
\begin{eqnarray}
 ( A \leftarrow  \rightarrow T ) \quad \Longrightarrow \quad \label{corr2}
 ( A_{imino} \leftarrow  \rightarrow T_{enol} )
 \\
 ( G \leftarrow  \rightarrow C ) \quad \Longrightarrow \quad \nonumber
 ( G_{enol} \leftarrow  \rightarrow C_{imino} ) \nonumber
\end{eqnarray}
During the replication, tautomeric transition driven by the proton
tunneling in conjunction with the complimentarity according to (\ref{corr1})
may lead to the change of base-pairs
\begin{eqnarray}
 ( A \leftarrow  \rightarrow T ) \quad \Longrightarrow \quad \label{corr3}
 ( G \leftarrow  \rightarrow C )
 \\
 ( G \leftarrow  \rightarrow C ) \quad \Longrightarrow \quad \nonumber
 ( A \leftarrow  \rightarrow T ) \nonumber
\end{eqnarray}
and result in loss, or corruption, of genetic information, i.e.
mutations.\cite{CW2,Loew}  The specific case given by the diagram
(\ref{corr3}) is called transition mutations; it has the property
of being reversible, i.e. able to go back to the wildlife type.

The arguments given above constitute the main points of the theory
of spontaneous mutations suggested by Crick and
Watson.\cite{CW2,Loew,Crick,topal-fresco} It is based on the
assumption that the transitory tautomeric shifts of base-pairs may
occur during the replication, i.e.  when two molecules of DNA are
formed from a paired molecule, so that the double-stranded
molecule is split into two single strands, each of which  controls
the synthesis of a new strand complimentary to itself with the
help of the special enzyme called DNA polymerase. It has been
realized that the latter plays an active role in the selection of
bases at replication\cite{Auer}, so that it may affect the
mutation rates.  Thus, tautomeric transitions are not a unique
cause of mutation; the situation is more subtle, and many
questions, of quite a classical nature, wait their solutions.
Nonetheless, the original idea of Watson and Crick still conserves
its appeal, and even more so as its new links with other phenomena
related to the mutagenesis are brought to light
(see\cite{Topal,Robinson}). So, Robinson et al\cite{Robinson},
report that the enol tautomer of $iG$, that is
$2^{\prime}-$deoxyisogine, may form at physiological temperature
($37^o$) and pair with thymine in a Watson-Crick geometry; thus,
$iG$ being present as the nucleoside, results in the formation of
incorrect base-pairs during in vitro replication.
\cite{K1,K2,K3,S1,S2} Robinson et al\cite{Robinson}, suggests that
$iG \cdot T$ pairing may have a bearing on mutagenesis in vivo
involving tautomers of the common nucleobases. On the other hand,
Fresco et al\cite{suen}, have found that the imino tautomer
$HO^5dCyt$ may serve as an example of an unfavored base tautomer
making for substituting mutagenesis.

Mutations within the framework of the Crick-Watson model of DNA
and in conjunction with the concept of tautomeric transition, have
been drawing attention,  beginning from the early fifties
\cite{CW1,CW2,Loew,Crick,topal-fresco}, to the present time, and
involved the use of condensed matter theory. So, one of the first
papers in this direction was published by Geracitano and
Persico\cite{Persico}, who suggested that there should be expected
a collective behavior of codons, resembling that taking place in
hydrogen-bonded ferroelectric crystals.

In this paper we intend to look after the interplay between
tautomeric transitions caused by the proton tunneling in
base-pairs ( see FIG.~\ref{fig1}) and elastic properties of the
double helix.

We feel that quantum effects caused by the proton tunneling may
have an appreciable bearing on mutagenesis.  In this respect we
would like to draw attention to the fact that  mutations could be
generated by irradiation with  electromagnetic waves  in infra-red
region corresponding to the energies of tautomeric shifts in the
base pairs of DNA. In fact, as was pointed out by Sukhorukov et
al\cite{Sukhorukov}, the available data on the absorbtion spectra
at $\nu=1697 cm^{-1}$, for the synthetic polynucleotide
(PolyU)-(PolyA) forming a two-stranded structure, may indicate the
transfer of protons between purine and pyrimidine
bases.\cite{Kyogoku} In papers\cite{Sukhorukov,Sukh} there is
reported an absorbtion band at $\nu=1712 cm^{-1}$, for DNA at
certain values of pH; the authors claim that the effect could be
due to the proton tunnelling in hydrogen bonds of base pairs.
Similar results are obtained in\cite{SSSM} for two-stranded
(PolyC) in LB-films.

\begin{figure}
  \begin{center}
     \includegraphics*[width=8.5cm, height=1.860cm]{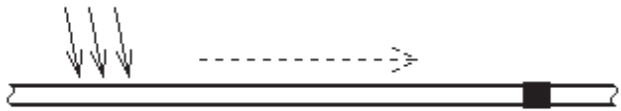}
     \caption{\label{fig24}Mutation (black square) in a region different
     from a set of base pairs subject to an initial
     mutagenic action (arrows).
     \label{fig2}}
  \end{center}
 \end{figure}

It is worth noting that the interplay among the proton tunneling
and the elastic properties of DNA may manifest itself in the
dynamical properties of mutagenesis. To be specific, it could
result in the phenomenon that the action imposed upon a set of
base pairs of DNA may finally cause a substitution mutation in a
different region of base pairs. One could expect the effect
similar to the freak waves of nonlinear theory, when a low
intensity initial perturbation for which the probability of
mutation per base pair is low, could focus on a few base pairs and
result in a mutation, ( see FIG.~\ref{fig2}).

The main point about our hypothesis is that, since the
$\pi-$electrons of the tautomeric rings of the nucleotides have
direct bearing on the interaction of the plates of adjacent
base-pairs\cite{Chris,HunSand}, a tautomeric transition of
base-pairs should substantially influence the distribution of
delocalized electrons of the nucleotides, i.e. the
$\pi-$electrons, and result in deformation of the elastic system
of DNA. The hypothesis is in accordance with the conclusions of
paper.\cite{Sukh} It is worth noting that tautomeric transitions
may occur in several base pairs, not necessary adjacent, at a
time, and their dynamics is determined by the proton tunneling. In
fact, for one thing the latter is due to electrostatic
interaction, i.e. the dipole forces, between the protons belonging
to adjacent base-pairs, and for another  the elastic system of the
DNA molecule, which  plays a role like that of the crystalline
lattice of the polaron theory. According to D.~Landau's original
idea, a charge moving in the crystalline lattice, deforms the
latter so that an effective field $U_{eff}$ is generated, and
within the framework of a self-consisted picture its motion is
determined by $U_{eff}$. This argument can be also applied to the
propagation of excitations of molecules constituting the lattice,
that is the exciton theory. An important specific case is the
regime in which the lattice relaxes to an equilibrium state fast
enough so that we could neglect its motion and consider only the
motion of the exciton, or charge. Then we may write down an
effective hamiltonian that allows for the deformation of the
lattice caused by the exciton, or charge, and next using an
appropriate trial function derive an equation for its
motion.\cite{Holstein} Davydov\cite{Davy}, used the idea for the
theory of $\alpha$-helix in proteins, the molecule of protein
playing the part of the crystalline lattice, and
Volkenstein\cite{Vol}, for his conformon theory.

The essential point about the possible interplay among the proton
tunneling and the conformation of the DNA is the values of its
elastic constants. In fact, as far as the transport of torsional
stress (torque) along DNA is concerned, its estimates obtained by
various means diverge widely. The numerical values derived with
the help of the theory of continuous media\cite{LeviCr}, are of
the order $\tau \propto 10^{-17} dyne \cdot cm
$\cite{Nelson,LiuWang}, whereas the experimental
evidence\cite{ma}, indicates that it can attain the value of $\tau
\propto 10^{-13} dyne \cdot cm$. Philip Nelson\cite{Nelson},
suggested that these deviations could be due to small bends  in
the helix backbone, so that one may assume
$$
   \tau \propto 10^{-17} \div 10^{-13} \quad dyne \cdot cm
$$
For describing the elastic properties of the double-helix we may
use the approach worked out in \cite{Peyr}, \cite{99:b},
\cite{gv}. Thus, the double helix is considered as a
one-dimensional lattice of vectors $\vec y_n$ describing the
mutual position of the two strands at sites corresponding to the
base-pair of index n.  It is important that the system has a
twisted ground state characterized by the twist vector $\Omega$,
so that the elastic energy of the molecule can be cast, at least
for sufficiently small $\vec y_n$, in the form
\begin{equation}
    H_{tor} =
     \sum_{i=1}^N \left[ \
        \frac{1}{2} M \left ( \partial_t \vec y_i  \right )^2
     +
        \frac{1}{2} K \left ( \nabla \vec y_i  \right )^2
     +
        \frac{1}{2} \epsilon \, \vec y_i \,^2
    \right ]    \label{Ten}
\end{equation}
where the first term is the kinetic energy,
the second one the elastic torsional energy
and the last one corresponds to the separation of the two strands.
The covariant derivative that accommodates the torsion of the molecule, reads
$$
  \nabla \vec y_i = \frac{1}{a} \,
            \left( \vec y_{i+1} - \vec y_i
            + \vec \Omega \times \vec y_i
                    \right )
$$
Here $a$ is the spacing between the adjacent nucleotides, M is the
mass of base-pair. For the sake of simplicity,  we shall assume
that the torsion vector $\vec \Omega$ is always parallel to the
axis Oz, that is
$$
     \vec \Omega = (0,0, \Omega)
$$
and the vectors $\vec y_n$ describe only  transversal motions, that is $y_n^3 = 0$. It should be noted
that we consider a very simplified model, use the harmonic approximation for its elastic energy, and
assume that all sites, corresponding to base-pairs are identical. The subtle question is the value of
the elastic constant $K$; obviously enough it has a direct bearing on the torque $\tau$ mentioned
above, and therefore, as was discussed above, its estimate may read
$$
     K \propto 10^{-17} \div 10^{-13} \, erg
$$
It should be noted that the calculations within the framework of
molecular dynamics, (see paper\cite{lavery}  and references
therein), give the upper value for $K$, i.e. close to $10^{-12}
\div 10^{-13} \quad erg$.

The interplay between the torsional stress due to the relative
motion of the base-pairs and the proton tunneling is very
important. As was mentioned above the tautomeric transitions are
driven by the proton tunneling, and therefore we shall describe
them quantum mechanically, that is the stable amino/keto form
corresponding to the ground state of proton, and the unstable
imino/enol one to the excited state.\cite{Sch1} In accord with the
qualitative character of our approach we neglect the fact that the
tautomeric transitions in question, involve the tunneling of more
than one proton, and assign only one proton to each site of the
lattice. There are few hydrogen bonds in which the protons are
transferred  towards the imino/keto groups, or if one uses the
concept of the two-level system, excited states. Therefore, one
can consider the system as being close to equilibrium, or only
weakly excited.  This suggestion is very important for what
follows.

We shall describe the states of a base-pair at site $n$ with the Bose
operators $b_n^+,  \, b_n$ that verify the usual conditions
$$
    [ b_n, b_m^+] =  \delta_{nm}, \quad
    [ b_n, b_m]   = [ b_n^+, b_m^+] = 0,
$$
and accommodate the assumption that the tunneling states of
protons be described as two level systems by {\it considering only
their ground states and the first excited state}. The energy of
the protons, neglecting the interaction with the elastic degrees
of freedom, reads\cite{Sch1}
$$
    H_P= \sum_n E_o b_n^+ b_n \quad + \quad
             \kappa \sum_n (b_n^+ b_{n+1} + b_{n+1}^+ b_n)
$$
Here $E_o$ is the energy of the tautomeric shift; its estimates depend on the choice of nucleotide and
according to quantum chemistry calculations vary within the range of $2 \div 10 \, Kcal$,
(see\cite{Saenger} and references therein). The constant $\kappa$ could be ascribed to dipole
interactions between adjacent sites, similarly to Davydov's theory.\cite{Davy} Presently, there are no
reliable estimates of its value (see below); by analogy with the Davydov theory one may assume that it
should correspond to the characteristic frequency of tautomeric excitation due to the proton
tunneling, that is of the order $10^{11} \; Hz$, or less. This estimate is generally accepted (see
below).

The central point of the model introduced in\cite{Sch1} is the
interaction between the elastic degrees of freedom of DNA and the
tautomeric transitions, or the proton tunneling in nucleotides; it
reads
$$
    H_I = - \lambda \, \sum_n \left( \nabla \, \vec y_n \cdot \vec h_n \right)
                    b_n^+ b_n
$$
Here the vectors $\vec h_n$ give the spatial orientation for the hydrogen
bonds of base-pairs
$$
   \vec h_n = ( \, \cos n\alpha, \, \sin n\alpha, \, 0 )
$$
It is important that the angle $\alpha$ is the rotation angle of the
double-helix, and thus close to the angle $\Omega$, in accord with
the fact that the covariant derivative term in the elastic energy
provides only a qualitative description for the DNA-helix.

An argument  in favor of the choice for the interaction $H_I$ is
that it takes into account the deformation of positions of
adjacent base-pairs and its influence on the $\pi-$electrons of
the bases, and therefore, the tautomeric transitions, or the
related excitations of protons. According to the theory
of\cite{HunSand}, the interaction could be appreciable. Thus, one
may suggest that the interaction term could be larger than the
tunneling term in the equation for $H_P$ given above.

Concluding we write  the total energy within the framework of the
model introduced in\cite{Sch1}  in the form

$$
    H_{total} = H_{tor} \quad + \quad H_P \quad + \quad H_I
$$

To find weakly excited states we shall use the Davydov
approximation\cite{Davy,AScot2}, that is we shall look for the
state vector of the system using the trial function
$$
     | {\cal D} > = \sum_n \, A_n(t) \cdot b^+_n \, |0 >
$$
where $|0>$ is  the ground state of the system, for which
all the base-pairs, or the protons in the hydrogen bonds, are in the ground
state, the amplitudes $A_n(t)$  being subject to the constraint
$$
    \sum_n \, |A_n(t)|^2 \, = 1
$$

The adiabatic approximation, which is important for the
implementation of the Davydov theory, holds for the following
reasons. The vectors $\vec y_n$ describe the dynamics of
base-pairs, that is relatively massive objects, and therefore one
may consider them as classical fields.\cite{Sch1,99:b}. We can
derive the size of characteristic frequencies for $\vec y_n$ from
expression (\ref{Ten}) of the elastic energy. In  fact, the mass
$M$ is that of the base-pair, that is of the order $500$ Dalton,
and $K$ is of the same order of magnitude as the torque $\tau$
discussed above. Hence, we get the characteristic velocity $v$ for
the $\vec y$ modes
$$
            v \propto \sqrt{ \frac{K}{M} }
$$
Interesting numerical values for the velocity $v$ follow from the equation
indicated above and the rough estimates for $\tau$ or $K$ we have mentioned.
Indeed, for $K \propto 10^{-17} dyn \cdot cm$ or less  we obtain
$$
            v \propto 10^2 \, cm/sec
$$
For wavelengths of a few tens of  $\AA$  it gives the
characteristic torsion or phonon frequencies of the order
$$
    \nu_y \propto 10^8 \div 10^9 \: Hz
$$
On the other hand, if we use the values for $K$ provided by the
molecular dynamics simulations\cite{lavery}, we get the velocity
of excitations of the order $1000 m/sec$, and $\nu_y \propto
10^{11} \div 10^{12} \, Hz$, as for  ordinary condensed media.

The elasticity of the DNA strongly depends on nucleotide sequence,
and therefore the arguments given above are only of qualitative
nature. Coleman et al, \cite{olson}, put forward a lattice model
of the DNA in which they try to accommodate the sequence
dependence of elastic properties. At each side the deformation of
the lattice is described by six kinematical variables: the three
angular variables $\theta_i^n$ (tilt, roll, twist) and the three
displacement variables $\rho_i^n$ (shift, slide, rise).  The
elastic energy $\Psi$ of a DNA segment is assumed to be the sum
$$ \Psi = \sum_n  \psi^n $$
of the interaction energies of adjacent base pairs $\psi_n$, which
are functions of the above kinematic variables. In the notations
of paper \cite{olson} energy $\psi^n$ reads
$$
    \psi^n  =  \frac{1}{2} \, F^n_{ij}(\Delta \theta^n_i)(\Delta \theta^n_j) +
                              G^n_{ij}(\Delta \theta^n_i)(\Delta \rho^n_j) +
               \frac{1}{2} \, H^n_{ij}(\Delta \rho^n_i)(\Delta \rho^n_j)
$$
where $F^n_{ij}, \ ,G^n_{ij}, \, H^n_{ij}$ are constants with
$F^n_{ij} = F^n_{ji}, \quad H^n_{ij} = H^n_{ji}, \quad i,j =
1,2,3$.  Coleman et al, \cite{olson}, estimate
$$
         F_{11} = F_{22} = 4.27 \times 10^{-2} \, \frac{k_B T}{deg^2}
$$
For $T = 300$ the above estimate gives $1.76 \times 10^{-14} \,
erg$, and the velocity of 'angular' waves corresponding to the
above expression for the energy, several hundred $m / sec$. If the
corresponding excitations of the double helix are of wavelength
comparable with the distance between adjacent base pairs, we get
the GHz-frequency range.  It should be noted that inter-strand
modes, which are likely to correspond to the angular waves in the
double helix, have been detected and measured in the region of
tens - hundreds GHz, \cite{urabe, urabe2,tao, tao2}. Consequently,
their propagation velocity could be well within a few hundred
m/sec.

It is instructive to compare the values of $\nu_y$ with the
transition frequencies for tautomeric reactions inside the
nucleotides,
$$
\nu_P = \frac{\kappa}{2 \pi \hbar}
$$
The estimates for the latter differ considerably,  \cite{Bodor},
$$
     \nu_P \propto 10^6 \div 10^{11} \, Hz
$$
The lowest estimate, $10^6$ Hz appears to be  not unreasonable
(V. Benderskii, and J.L.Leroy, personal communications).

The relative sizes of $\nu_P$ and $\nu_y$ are important
for choosing the right approximation for the model.
In fact, if we are at the lowest end of the spectra $\nu_P$,
then according to the estimate for $\nu_y$
obtained above the characteristic times for the acoustic modes are at
least by an order of magnitude smaller than for the protons. In this case,
we may suggest that the elastic system should follow the motion of the
protons in hydrogen bonds, adjusting itself to it, so that
a kind of adiabatic approximation can be employed.
In this paper we shall follow this conjecture.

Thus, we assume, as in paper\cite{Sch1}, that the adiabatic
approximation is valid, and therefore  neglect the kinetic energy
of the elastic system and take into account only its potential
energy generated  by the field $\vec y_n$. Then we are in a
position to apply the self-consisted method of the exciton theory,
in the form suggested by Davydov\cite{Davy}, that
is to calculate the mean value
\begin{equation}
        U_{eff} = < {\cal D}| H_{tor} + H_I |{\cal D}> \label{mean}
\end{equation}
find the minimum, $\vec y_n^{(o)}$ of $U_{eff}$  with respect to
$\vec y_n$, substitute it into the equation for the total energy
$H_{total}$ so as to get the effective Davydov hamiltonian ${\cal
H}_D$, which depends only on the operator variables $b^+_n, \,
b_n$, the classical variables $\vec y_n$ having disappeared
through the minimization. To make this paper as self consistent as
possible we reproduce the necessary calculations of
paper\cite{Sch1}. Thus, we obtain an equation that
has the form of the Schr\"odinger one
\begin{equation}
    i \hbar \frac{\partial}{\partial t} \, | {\cal D}>  =
                     {\cal H}_D | {\cal D} > \label{SD}
\end{equation}
and in which the wave function $| {\cal D} >$ should be of the
form prescribed above. The assumption that the excited states
correspond to the set of two-level systems is accommodated by the
requirement that the operators $b^+_n$ are allowed only in the
first power. It results in a system of equations, called the
Davydov equations, for the amplitudes $A_n$, which one obtains on
equating the coefficients at $b^+_n$ on both sides of (\ref{SD}),
( see\cite{Davy} for the details ). In this paper we consider the
case of the stretching energy for the DNA strands being smaller
than the torsional one, that is we assume
$$
   \frac{\epsilon a^2}{K \Omega^2}
$$
being small enough.

The Davydov hamiltonian for our problem reads
\begin{eqnarray}
  H_D &=& \sum_n E_0 b_n^{+}b_n
       -\sum_n \kappa (b_{n+1}^{+}b_n + b_n^{+}b_{n+1}) \label{Dham} \\
   & & -\frac{\lambda^2}{K} \sum_n|A_n|^4
       -\frac{\lambda^2}{K} \sum_n|A_n|^2 b_n^{+} b_n \nonumber \\
   & & +\frac{\lambda^2}{2K} \frac{\epsilon a^2}{K\Omega^2}
     \sum_{m,n}cos^{|m-n|}\phi \cdot cos \left[ (m-n)(\phi-\alpha) \right]
     |A_m|^2 |A_n|^2 \nonumber \\
   & & +\frac{\lambda^2}{2K} \frac{\epsilon a^2}{K\Omega^2}
     \sum_{m,n}cos^{|m-n|}\phi \cdot cos \left[ (m-n)(\phi-\alpha) \right]
     |A_n|^2 b_m^{+} b_m \nonumber
\end{eqnarray}

\noindent
and the equation for the amplitudes $A_n$

\begin{eqnarray}
  i \hbar \frac{\partial}{\partial t} A_n &=&
   E_0 A_n
   -\kappa(A_{n+1} + A_{n-1})
    -\frac{\lambda^2}{K}|A_n|^2A_n
       -\frac{\lambda^2}{K}(\sum_{m}|A_m|^4)A_n   \label{SchNL} \\
    &+& \frac{\lambda^2}{K} \frac{\epsilon a^2}{K\Omega^2}
     ( \sum_{m_1,m_2}cos^{|m_1-m_2|}\phi
     \cdot cos \left[ (m_1-m_2)(\phi-\alpha) \right]
     |A_{m_1}|^2 |A_{m_2}|^2 ) A_n  \nonumber \\
    &+&\frac{\lambda^2}{K} \frac{\epsilon a^2}{K\Omega^2}
     ( \sum_{m}cos^{|m-n|}\phi
     \cdot cos \left[ (m-n)(\phi-\alpha) \right]
     |A_{m}|^2 ) A_n                \nonumber
\end{eqnarray}

which has the form of a nonlinear discrete Schr\"odinger equation
for the amplitudes $A_n$.  The terms given by the third and the
fourth lines of the equation written above, describe an
interaction that has a very long range, so that one may claim
Eq.(\ref{SchNL}) is essentially nonlocal in character.

We shall make an important approximation that agrees with the general
qualitative setting of our work, and set
$$
    \alpha = \phi
$$
where $\phi = \arctan \, \Omega$. Thus, the
oscillating factors in Eq.(\ref{SchNL}) are cancelled out. With
the help of the reduced variables $B_n$
$$
    A_n = exp(-\frac{i}{\hbar} E_0 t) \, B_n(t)
$$
we cast the equation for $A_n$ in the form

\begin{eqnarray}
  & &i \hbar \frac{\partial}{\partial t} B_n =
   -\kappa(B_{n+1} + B_{n-1})
   -\frac{\lambda^2}{K}|B_n|^2B_n  \label{Sch2NL} \\
   & & -\frac{\lambda^2}{K}(\sum_{m}|B_m|^4)B_n \nonumber \\
   & & +\frac{\lambda^2}{K} \frac{\epsilon a^2}{K\Omega^2}
     ( \sum_{m_1,m_2}cos^{|m_1-m_2|}\phi |B_{m_1}|^2 |B_{m_2}|^2 )
     B_n  \nonumber \\
   & & +\frac{\lambda^2}{K} \frac{\epsilon a^2}{K\Omega^2}
     ( \sum_{m}cos^{|m-n|}\phi |B_{m}|^2 )B_n \nonumber
\end{eqnarray}

Introduce the characteristic frequencies
\begin{equation}
  \nu_P = \frac{\kappa}{2 \pi \hbar}, \quad
  \nu_T = \frac{\lambda}{2 \pi \hbar}, \quad
  \nu_{tor} = \frac{K}{2 \pi \hbar}
  \label{Freq}
\end{equation}
and the dimensionless time
$$
   \Upsilon = t \cdot \nu_P
$$
It should be noted that the frequencies $\nu_y$ and $\nu_{tor}$
are not identical, $\nu_y \neq\nu_{tor}$.
Then the Davydov equation takes the form
\begin{eqnarray}
  i \frac{\partial}{\partial \Upsilon} B_n &=&
   -(B_{n+1} + B_{n-1})
   -W|B_n|^2B_n \label{schnl} \\
   & & -W (\sum_{m}|B_m|^4)B_n \nonumber \\
   & & +W \Lambda
     ( \sum_{m_1,m_2}cos^{|m_1-m_2|}\phi |B_{m_1}|^2 |B_{m_2}|^2 )
     B_n \nonumber \\
   & & +W \Lambda
     ( \sum_{m}cos^{|m-n|}\phi |B_{m}|^2 )B_n \nonumber
\end{eqnarray}
in which
\begin{eqnarray}
  W &=& \frac{\nu_T^2}{\nu_P \cdot \nu_{tor}} \label{W}\\
  \Lambda &=& \frac{\epsilon a^2}{K \cdot \Omega^2}     \label{Lambda}
\end{eqnarray}

Now we aim at making the numerical simulation of Eq.(\ref{schnl})
for various values of the parameters $W, \, \Lambda$, looking for solutions
of the soliton type.  We use the term soliton in a sense close to that used by
applied scientists, i.e. a solution different from zero in a finite region
of space, whose {\it size we shall call the size of soliton},
and preserving its shape for very long periods of time.
For some values of $W, \, \Lambda$ it has the form identical to the
usual one, i.e. corresponding to the non-linear Schr\"odinger equation,
but generally our solitons are different.  The standard definition
suggests that it be of the form
\begin{equation}
 Y(x,t) = e^{i(qx - \nu t)}\, \psi(x -  vt)  \label{sts}
\end{equation}
in which $\psi$ is a real function.  It is by no means clear that
our solitons  always have the form given by Eq.(\ref{sts}).

\begin{figure}
  \begin{center}
     \includegraphics*[width=13cm, height=17.063cm]{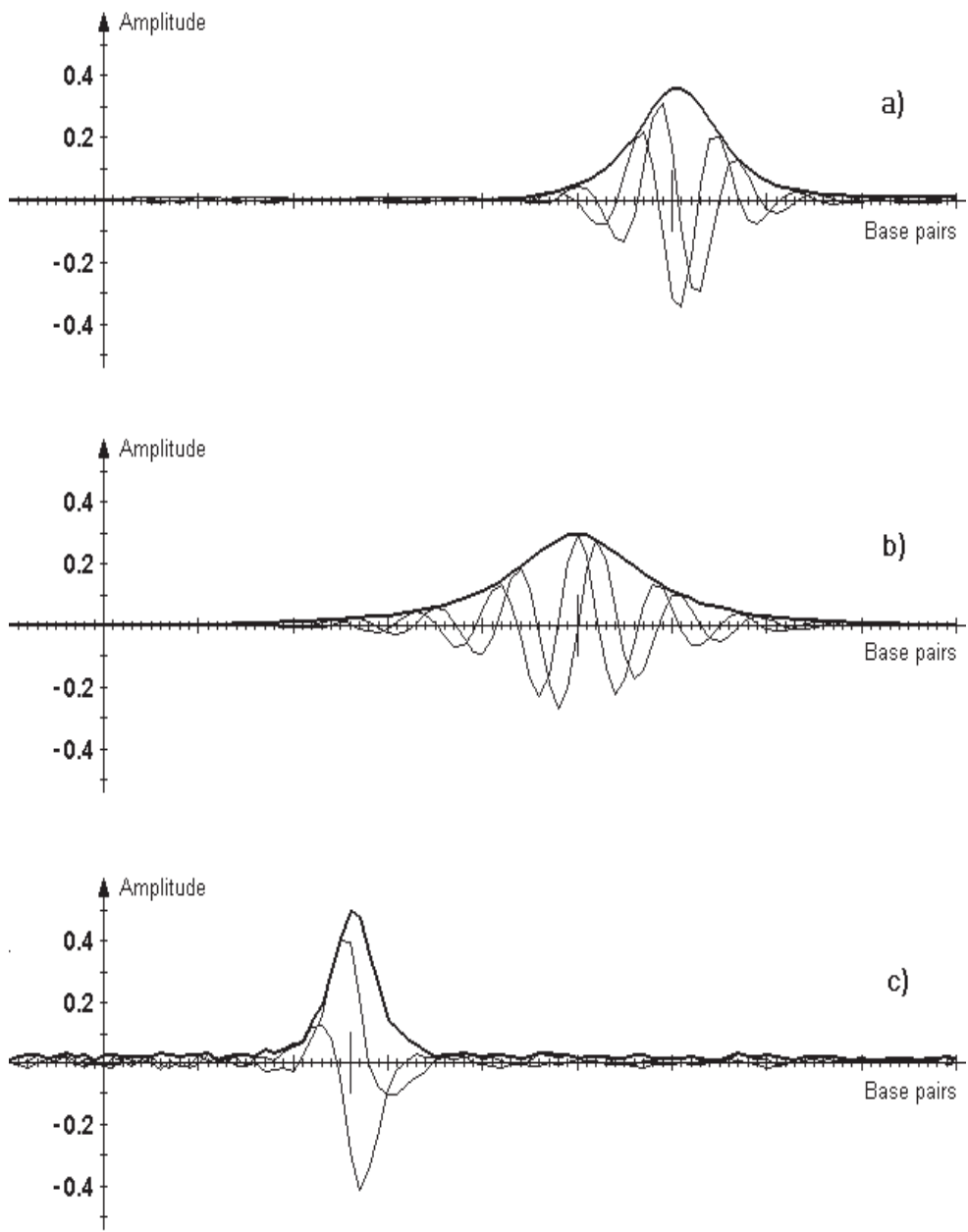}
     \caption{Typical moving solitons.
 Solid line:  $|A_n|$. Thin line: the real and the imaginary part
 of the amplitude $A_n$, $\nu_P = 10^6 \, Hz$.
 (a) for $W=0.75, \quad \Lambda= 0.001$,
     velocity $1330$ base pairs per $msec$, the period of time spent
     $0.533 \, msec$.
 (b) Moving soliton for $W=0.75, \quad \Lambda= 0.075$,
     velocity $1340$ base pairs per $msec$, the period of time spent
     $0.530 \, msec$, distance travelled $707$ base pairs.
 (c) Moving soliton for $W=2, \quad \Lambda= 0.1$,
     velocity $850$ base pairs per $msec$, the period of time spent
     $0.577 \, msec$, distance travelled $491$ base pairs.
     The values of $W, \, \Lambda$ are close
     to the borderline [see FIG.~\ref{diagram}~(b)],
     dividing the region of stable solitons from the unstable ones.
     \label{moving_solitons}}
  \end{center}
 \end{figure}

The parameter $\Lambda$ is a quantitative characteristic that
enables us to take into account the structure of the double helix,
and also the relative size of the torsional and deformation
energies. In fact, $\Lambda$ determines the magnitude of the
nonlocal terms in Eq.(\ref{schnl}), and in this respect it is
worthwhile to note that for certain values of $\Lambda$ and W we
have not been able to find soliton solutions, e.g. $\Lambda = 0.2$
and $W = 2$, at least for physically reasonable sizes of solitons,
i.e. less than 150 base pairs. The last constraint is due to the
fact that we consider straight segments of DNA, parallel to
Oz-axis, and therefore they should be of a size less than the
persistence length, that is about 150 base pairs. But it is
important that generally the condition $\Lambda \neq 0$ does not
forbid the existence of solitons, and its influence only results
in the size of soliton becoming larger, which is quite natural,
for $\Lambda$ represents non-local terms in Eq.(\ref{schnl}). The
general case of soliton with $\Lambda$ not equal to zero, even
though small, is illustrated in FIG.~\ref{moving_solitons}.
To understand the general situation let us consider the two
special cases.

\begin{figure}
  \begin{center}
     \includegraphics*[width=13cm, height=14.219cm]{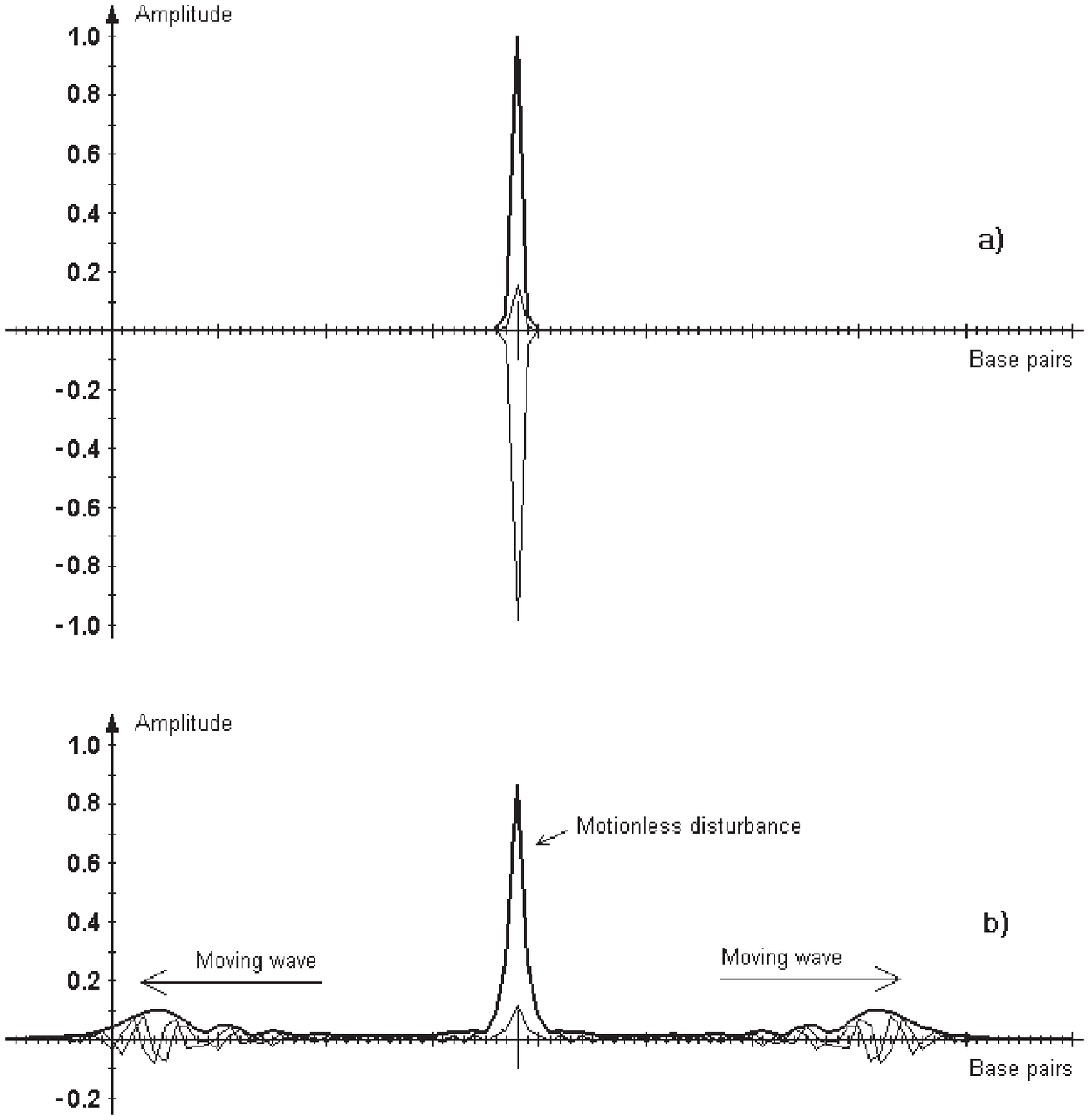}
     \caption{Typical breathers. Solid line:  $|A_n|$. Thin lines
 indicate the real and the imaginary part
 of the amplitude $A_n$,  $\nu_P = 10^6 \, Hz$.
 (a) Breather, or still soliton.
     $W = 10, \Lambda= 0.5$.
     The period of time
     spent $0.501 \, msec$.
 (b) Radiation emitted from  the motionless central peak
     during the period of time $0.018$, for $W=5$ and $\Lambda=0.5$;
     the velocity of side waves $1830$ base pairs/msec,
     distance traveled $33$
     base pairs.
     \label{breather}}
  \end{center}
 \end{figure}

1. Stationary solutions in the sense that the absolute value,
$|B_n(t)|$ does not depend on time. For the usual solitons given
by Eq.~(\ref{schnl}) this requirement means that the velocity $v =
0$. The typical case is illustrated in FIG.~\ref{breather}~(a),
for $W = 10$ and $\Lambda = 0.5$. The half-width of soliton equals
to one spacing between base-pairs, that is the solution is
extremely narrow, and according to our main hypothesis it must
correspond to the tautomeric transition of a base-pair. The very
interesting case is illustrated in FIG.~\ref{breather}~(b), $W =
5$ and $\Lambda = 0.5$. There is a central peak of half-width $1.5
\cdot a$ which stands still, and two symmetrical wave packets,
moving in opposite outward directions. The distance traveled by
these wave packets during 0.018 msec is equal to 33 base pairs.
The value of $\nu_P$ was taken $10^6 \, Hz$.

2. The usual solitons given by Eq.(\ref{sts}). The half-width of
these solitons may be several tens of base-pair spacings, and thus
they could correspond to tautomeric transitions taking place in
adjacent base-pairs. The typical cases are illustrated in
FIG.~\ref{moving_solitons}. It is interesting to note that these
solitons move, even though slowly. Their velocity is given by the
asymptotic formula
$$
  v \approx 2 \, a \nu_P \sin(a q)
$$
Hence, one might suggest the picture of  tautomeric
transitions moving along the DNA-molecule.

Both types of solutions indicated above are stable
with respect to perturbation of W and $\Lambda$.

Perhaps, the most characteristic feature of discrete non-linear
Schr\"odinger equation is solutions that periodically oscillate in
time and decay exponentially in space, or breathers. From a purely
qualitative point of view the existence of breathers can be
inferred from a truncated version of Eq.(\ref{schnl}). Let us
neglect all the terms on its RHS except the first two, that is
consider

$$
  i \hbar \frac{\partial B_n}{\partial t} =
  - (B_{n+1} + B_{n-1})  - W |B_n|^2 B_n
$$

\noindent
and look for $B_n$ such that

$$
   B_n = e^{i \nu t} a_n
$$

\noindent
$a_n$ being real.  Next, cast the equation for $a_n$ in the form

$$
   - (a_{n+1} - 2 a_n + a_{n-1})  -  [W a_n^3  + (2 - \epsilon)] a_n = 0
$$

Suppose that the soliton we are looking for is large enough so that
we may change the expression  $a_{n+1} - 2 a_n + a_{n-1}$ for the second
derivative.  Thus we obtain the equation

$$
  a^{\prime \prime} + [W a^3  + (2 - \epsilon)] a  = 0
$$

\noindent
or the conservation law for one dimensional motion with the
effective potential

$$
   V = \frac{2 - \epsilon}{2} a^2  \, + \, \frac{W}{4}  a^4
$$

\noindent
The soliton solution exists for $\epsilon \geq 2$, and its size
tends to infinity as $\epsilon \rightarrow 2$.  On the other hand
for large $W$ we may expect thin solitons.

\begin{figure}
  \begin{center}
     \includegraphics*[width=13cm, height=11.375cm]{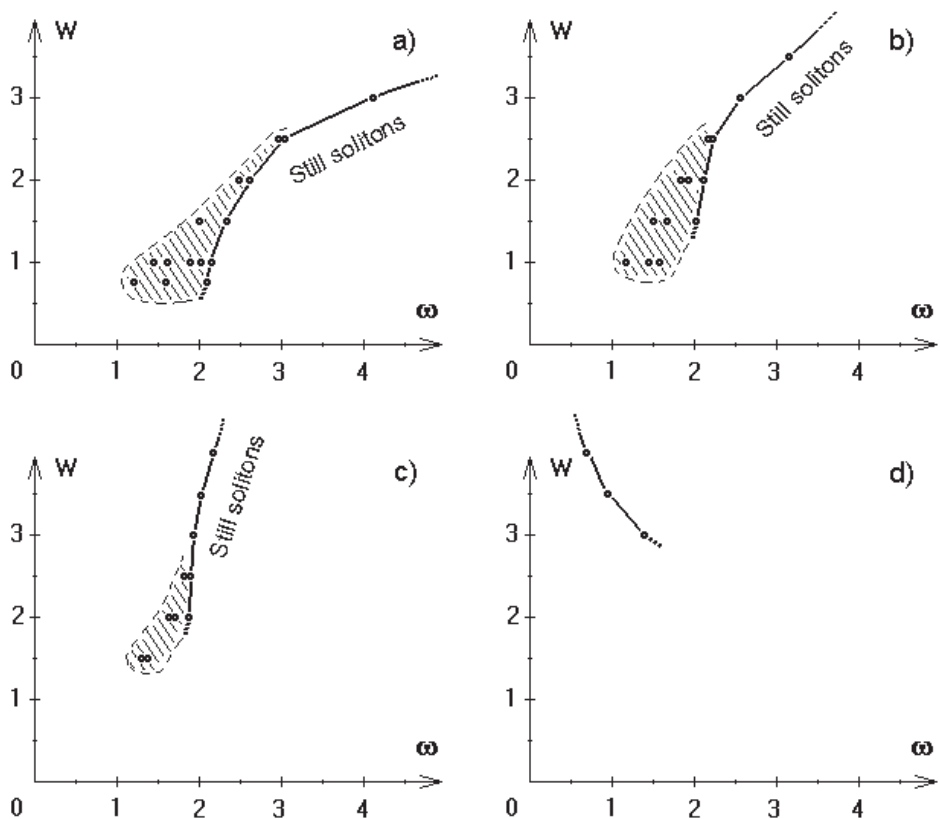}
     \caption{Sets of $W$ and $\nu$ that allow for
    soliton solutions. Solid line represents still solitons, or
    breathers. Shaded area represents moving solitons. Points are
    trial solutions. The transition from solitons to breathers goes
    smoothly as the soliton speed decreases to zero. We take into
    account only solutions of size less than 100 base pairs.
 (a)  $\Lambda = 0$ .
 (b)  $\Lambda = 0.1$.
 (c)  $\Lambda = 0.15$.
 (d)  $\Lambda = 0.2$.
 \label{diagram}}
  \end{center}
 \end{figure}

The key point is that the nonlocal terms generated by the double
helix bring serious modifications to the picture given above.
Thus, we may infer that the dimensionless constants $W$  and
$\Lambda$ play a crucial role in determining the form of solitons
for Eq.(\ref{schnl}).  The general situation to the effect is
illustrated in FIG.~\ref{diagram},
in which the horizontal axis corresponds to values of $\nu$, that
is the soliton frequency measured in units of $\nu_P$. It should
be noted that $\nu$ defines only the main Fourier component both
for solitons and for breathers, so that $\nu$ turns out to be only
a rough characteristic. The breathers are represented by the solid
line, that serves also as a right-handed border for the region of
moving solitons. This line continues up to infinity with both $W$
and $\nu$ rising. It's lower end was not clearly found. The lower
left-handed border of the soliton region is not strictly defined,
owing to the fact that there are solitons for values of $W$ and
$\nu$ lower than the borders but of sizes greater than $100$ base
pairs, that is beyond the physical context of our problem. The
upper left-handed part of the border is determined by solitons
turning out to be unstable for values of $W$ and $\nu$ beyond the
boundary, and thus is not defined clearly. We see that the soliton
region is decreasing as $\Lambda$ grows (see FIG.~\ref{diagram}),
and for $\Lambda = 0.2$, FIG.~\ref{diagram}~(d), there are only
breathers, at least under the constraint of their size being less
than $100$ base pairs. It is worth noting that Eq.(\ref{schnl})
derived in\cite{Sch1} is valid only for small $\Lambda$.

\begin{figure}
  \begin{center}
     \includegraphics*[width=13cm, height=9.141cm]{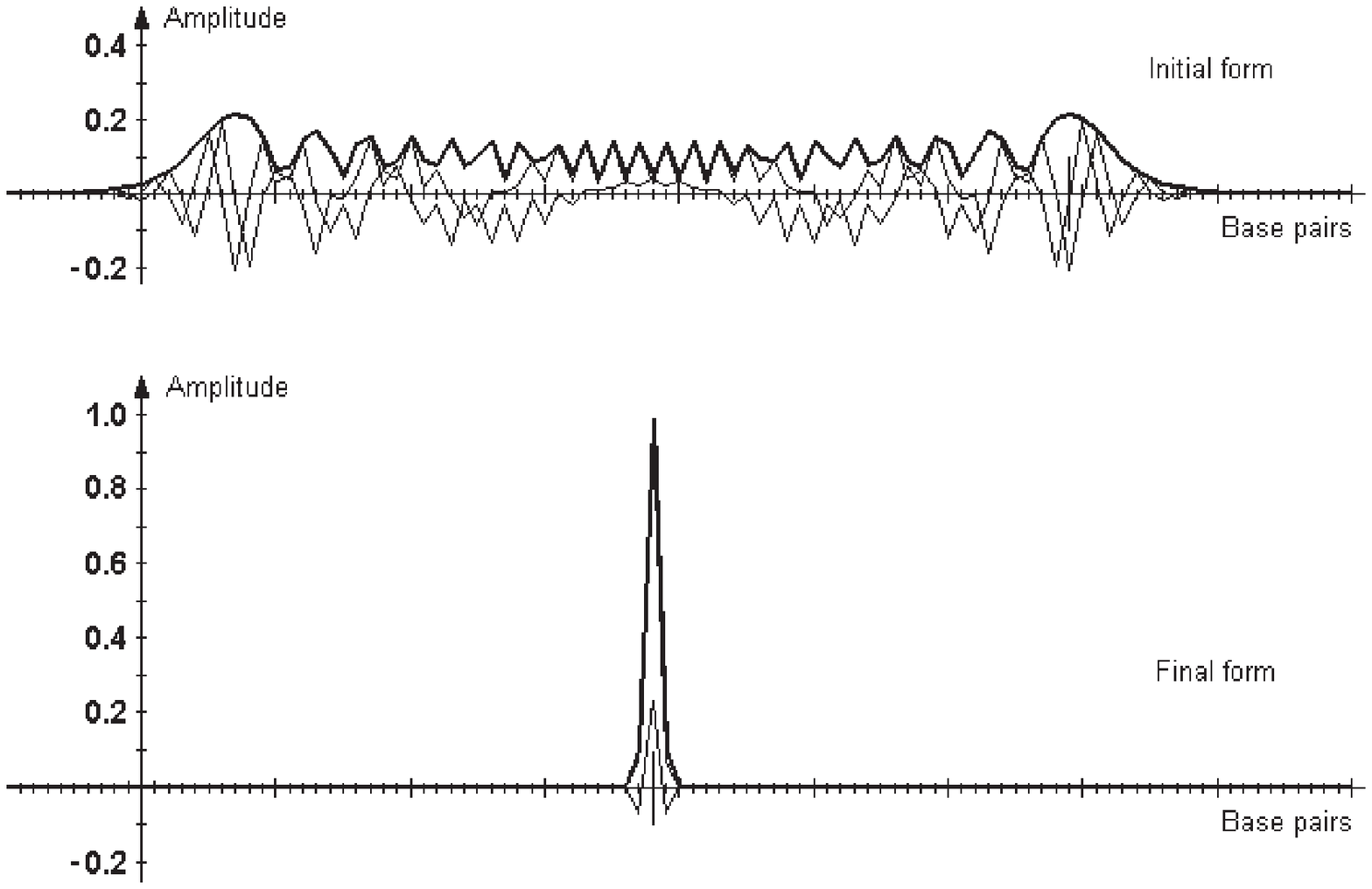}
     \caption{Partial self-focusing of an initial
     low amplitude distribution on a peak
 for a period of time $0.002 \, msec$, for
 $W=1, \, \Lambda = 0.5$, $\nu_P = 10^6 \, Hz$.
 \label{freak}}
  \end{center}
 \end{figure}

We would like to draw attention to a class of solutions that
correspond to the nomenclature of "freak
waves"\cite{slun1,slun2},and which  may have a bearing upon the
dynamics of tautomeric transitions. A solution of the type is
illustrated in FIG.~\ref{freak}.
It is characterized  by an initial set of amplitudes $B_n(t)$
which is a broad distribution of the size of 80 base-pair
spacings; after the period of time 0.017 msec, the characteristic
frequency $\nu_P$ being taken $10^6 \, Hz$, it focuses itself on a
narrow peak of half-width of one spacing. The peak exists for the
brief period of time 0.002 msec, and next breaks down into a broad
distribution again, i.e. a kind of partial self focusing is taking
place. Thus, there may exist low probability tautomeric
transitions distributed over wide areas of the molecule, and which
may collapse into a small region of the molecule, and stay there
for a period of time, brief but perhaps sufficient to cause
mutation. For finding the initial configurations producing the
peaks indicated above we  used the method of numerical integration
backward in time, similar to that used in papers.
\cite{slun1,slun2}

\begin{figure}
  \begin{center}
     \includegraphics*[width=13cm, height=7.008cm]{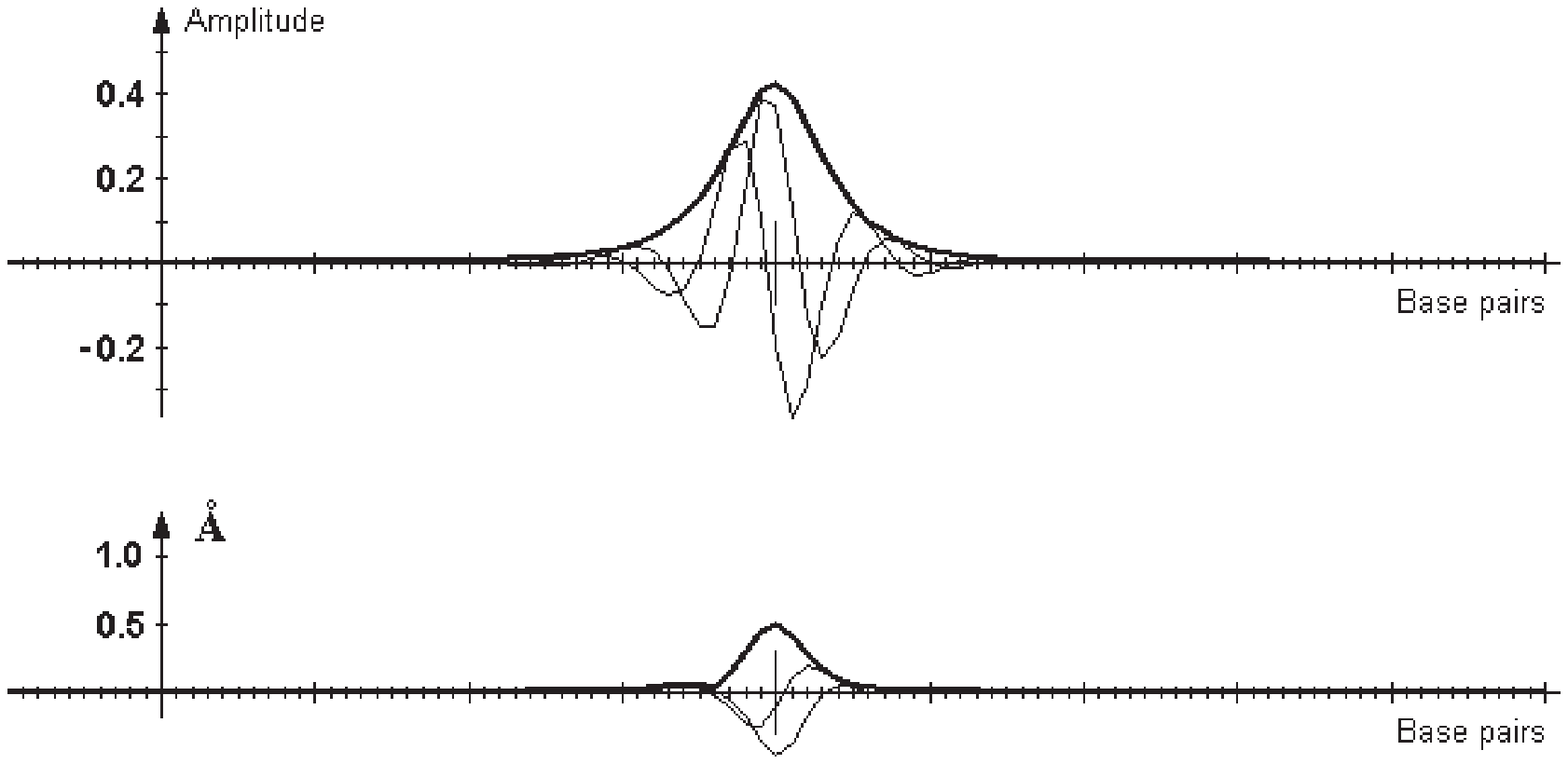}
     \caption{
        Moving soliton for $W=2 \quad \Lambda= 0.1$,
        velocity $880$ base pairs per $msec$, $\nu_P = 10^6 \, Hz$ .
        Maximum $|\vec{y}_n| = 0.52 \AA$
        (a) The amplitude $A_n$.
             Solid line:  $|A_n|$. Thin line: the real and the imaginary part
             of the amplitude $A_n$.
        (b) $\vec{y}_n$ distribution.
            Solid line:  $|\vec{y}_n|$. Thin line: the first and second
            coordinates of the $\vec{y}_n$.
      \label{proton_yn}}
  \end{center}
 \end{figure}

It is instructive to see the conformation of the field $\vec y_n$
accompanying the dynamics of solitons. The typical configuration
of $\vec y_n$ corresponding to the soliton solution of
Eq.(\ref{schnl}) is shown in FIG.~\ref{proton_yn}.

We see that the formation and the dynamics of solitons
corresponding to the tunneling of protons is accompanied by
conformational changes. Thus, we may infer that the tautomeric
transitions studied in this paper are to a large extent similar to
conformons.\cite{Vol} In fact, the concept of conformon was
suggested for describing the dynamics of charge transfer in
macromolecules, especially proteins, similarly to the situation we
are considering.

It is also interesting to invert the picture discussed above, and
suggest that there is a deformation of the double helix described
by a distribution of $\vec y_n$, like that shown in
FIG.~\ref{proton_yn}, then there should be a distribution of the
soliton amplitude, that is a proton tunneling generated by the
conformational change. Thus, we may suggest that the
conformational transitions may result in the proton tunneling, or
the tautomeric  shifts.

As was shown above, the dynamics of tautomeric transitions in DNA
depend on elastic properties of the latter and proton tunneling in
base pairs; $W$ and $\Lambda$ serving as indicators for possible
regimes. Our numerical simulation suggests that the interesting
tautomeric dynamics may happen for $W \ge 1$. This allows for
sufficiently wide range of material constants of DNA so that the
phenomenon could occur. The second constant, $\Lambda$, provides a
quantitative characteristic for the part played by the double
helix; it can totally modify the structure of solitons
corresponding to tautomeric transitions.

Depending on the value of W one may expect the existence of two quite
different dynamics: (1) solitons that move at velocities smaller
by orders of magnitude compared with that of elastic excitations in DNA,
and have a size of several tens of base-pairs,  and
(2) stationary solutions, or breathers, that have a form of peaks over
a few base-pairs. We may suggest that the second type of solutions
correspond to  point mutations, whereas the first one
may describe tautomeric transition moving along the chain of
double helix, and therefore there may happen mutations
related to the transition.
Thus, one may suggest that an action imposed on a set of nucleotide
in a region of the molecule might generate mutations in a different region
owing to the motion of excitations corresponding to the proton tunneling.

It is alleged to be known that by substituting the "artificial"
nucleotides instead of the natural ones, e.g. brom-uracil for thymine,
one can increase dramatically the rate of mutations;
this could be due to the increase of tautomeric transitions inside
base-pairs. At any rate, it is worthwhile to study the interplay between
the rate of such transitions and mutations. Within the context of the
present paper, artificial DNA of this kind could ease the stringent
constraints imposed on $W$, as was indicated above.

It is worth noting that the "focusing" of solutions (see FIG.~\ref{freak}), similar to the freak waves
which take place in the theory of non-linear waves\cite{slun1,slun2}, may have a very important
bearing on mutations. In fact, it amounts to the possibility of a weak external influence generating a
low amplitude distribution of mutation sites that would focus itself later on a high amplitude
distribution concentrated in a different region of the molecule. Thus, one may expect generating
mutations by low intensity agents distributed in a region of the molecule, or to put it the other way
round, acting on a set of codons different from those that suffer the actual mutation.\vspace{4cm}

\centerline{ACKNOWLEDGEMENT} I am thankful to the participants of the workshop (SCA) for the
attention.


\begin{thebibliography}{99}
    \bibitem{calladine}  M.A.El Hassan and C.R.Calladine  1995
            J.Mol.Biol. {\bf 251} 648
    \bibitem{Saenger} W.Saenger,
            Principles of Nucleic Acid Structure,
            Springer-Verlag, New York (1984).
    \bibitem{SantaLucia} John SantaLucia, Jr.,
            Proc.Natl.Sci.USA {\bf 95}, 1460 (1998).
   \bibitem{marko} J.F.Marko and E.D.Siggia,  Phys.Rev. {\bf 52}, 2912 (1995).
   \bibitem{sig1}  J.F. Marko and E.D. Siggia, \
            Macromolecules, {\bf 27}, 981 (1994).
   \bibitem{bloomfield}  C.G.Baumann, S.B.Smith, V.A.Bloomfiled,
                          and C.Bustamante,
            Proc.Nat.Acad.Sci. USA, {\bf 94}, 6185 (1997).
   \bibitem{bryant} Z. Bryant, M. D. Stone, J. Gore, S. B. Smith, N. R. Cozzarelli,
                    and C. Bustamante, Nature, London {\bf 424}, 338 (2003).
   \bibitem{georghiou} Chia C.Shih and S. Georghiou,
            J.Biomo.Str.Dyn. {\bf 17}, 921 (2000).
   \bibitem{kovaleva1} N.A. Kovaleva and L.I. Manevitch,
                      Localized nonlinear oscillation of DNA molecule,
                      Proceedings of 8th Conference on Dynamical Systems – Theory and Application,
                      Lodz, Poland, 2005, p.103.
   \bibitem{kovaleva2}N.A. Kovaleva, L.I. Manevitch, and V.V. Smirnov,
                      Analytical study of coarse-grained off-lattice model of DNA,
                      DSTA-2007, 9th International Conference: on Dynamical Systems :
                      Theory and Applications ,Lodz, Poland. P. 74.
   \bibitem{us}Thomas A. Knotts, Nitin Rathore, David C. Schwartz, and Juan J. de Pablo,
                      J. Chem. Phys. 126, 084901 (2007).
   \bibitem{urabe} H.Urabe, Y.Sugawara, M.Tsukakoshi, and T.Kasuya,
            J.Chem.Phys. {\bf 95} 5519 (1991).
   \bibitem{urabe2} H.Urabe and Y.Tominaga, Y. J.Phys.Soc.Jpn.
            {\bf 50} 3543 (1981).
   \bibitem{proh1} Y.Kim and E.W.Prohofsky 1987  Phys.Rev. {\bf B36} 3449.
   \bibitem{CW1} J.DWatson and  F.H.C.Crick,
            Nature    171, 737 (1953).
   \bibitem{CW2} F.H.C.Crick and J.D.Watson,
            Brookhaven Symp. Biol. No. 8 (1956).
   \bibitem{Loew} P.O.L\"owdin,
            Rev.Mod.Phys. {\bf 35}, 724  (1963).
   \bibitem{Crick} F.H.C.Crick,
                     J.Mol. Biol., {\bf 19}, 548 (1966).
   \bibitem{vinogr}  J.Vinograd, J.Lebowitz, R.Radolf, R.Watson,  and P.Lapis,
                     Proc.Natl.Acad. Sci. U.S.A. {\bf 53}, 1104 (1965).
   \bibitem{LL} L.D. Landau and E.M. Lifshitz,
                     Theory of Elasticity, Ch. 2, Moscow, "Nauka" (1987).
   \bibitem{nelson1} Randall D.Kamien, To.C.Lubensky, Philip Nelson,
                     and C.S.O'Hern,
                     Europhys.Lett. {\bf 38}, 237 (1997).
   \bibitem{nelson2} C.S.O'Hern, Randall D.Kamien, T.C.Lubensky,
                     and Philip Nelson,
                     Elasticity Rheory of a Twisted Stack of
                     Plates, cond-mat/9707040 vl 3Jul 1997.
   \bibitem{nelson3} Philip Nelson,
                     Proc.Natl.Acad. Sci. U.S.A. {\bf 96}, 14342 (1999).
   \bibitem{nelson4} J.David Moroz and Philip Nelson,
                     Proc.Natl.Acad. Sci. U.S.A. {\bf 94}, 14418 (1997).
   \bibitem{nelson5} J. David Moroz and Philip Nelson,
                     Macromolecules {\bf 31}, 6333 (1998).
   \bibitem{lavery} M.Bruant, D.Flatters, R.Lavery, and D.Genest,
                     Biophys. J. {\bf 77}, 2366  (1999).
   \bibitem{khokhlov} A.R.Khokholov and A.Grosberg,
                     Statistical Mechanics of Macromolecules,
                     Ch.1, AIP, New York (1994).
   \bibitem{tjablikov} S.V.Tjablikov, Methods of Quantum Theory of
                      Magnetism, (in Russian) Nauka , Moscow (1965).
   \bibitem{Stari} E.B.Starikov, Phys.Rep. {\bf 284}, 1(1997).
   \bibitem{smith}  S.B.Smith, Y.Cui, and C.Bustamante,
                     Science {\bf 271}, 795 (1966).
   \bibitem{tao} Tao N.J. and Lindsay S.M., Biopolymers {\bf 28}, 1019 (1989).
   \bibitem{tao2} Tao N.J., S.M.Lindsay, and A.Rupprecht,
                     Biopolymers, {\bf 27},  1655 (1988).
    \bibitem{gk1} V.L.Golo and E.I.Kats,
                     Pisma ZhETF {\bf 62}, 605 (1995) ).
   \bibitem{gk2} V.L.Golo and E.I.Kats,
                     ZhETF {\bf 108}, 1 (1995).
   \bibitem{feynman}  R.Feynman, Statistical Mechanics,
                     Ch. 2,Benjamin, Reading,
                     Massachusetts (1972).
   \bibitem{manning} G.S.Manning, Biopolymers, {\bf 22}, 689 (1983).
   \bibitem{Leroy1} M.Gu\'eron, M.Kochoyan, and J.L.Leroy,
                     Nature {\bf 328}, 89 (1987).
   \bibitem{Leroy2} J.L.Leroy, X.Gao, V.Misra, M.Gu\'eron,
                     and D.J.Patel, Biochemistry, {\bf 31}, 1407 (1992).
   \bibitem{Leroy3} M.Gu\'eron and J.L.Leroy, Methods Enzymol. {\bf 261}, 383 (1995).
   \bibitem{Bonnet} G.Altan-Bonnet, A.Libchaber, and O.Krichevsky,
                     Phys.Rev.Lett. {\bf 90}, 138101 (2003).
   \bibitem{kats}  V.L.Golo,  E.I.Kats and M.Peyrard,
                     Pisma ZhETF, {\bf 73}, 225 (2001).
   \bibitem{microwave} V.L. Golo, ZhETF {\bf } to appear (2005).
   \bibitem{zandt}   M.E.Davis and L.L.VanZandt,
                     Phys.Rev. {\bf A 37}, 888 (1987).
   \bibitem{Rayleigh}  J.W.Rayleigh,  1926,  The Theory of Sound,
                     Ch. III vol. I  (London: MacMillan)
   \bibitem{powell} J.W.Powell, G.S.Edwards, L.Genzel and A.Wittlin,
                     Phys.Rev. {\bf A35},  3929 (1987).
   \bibitem{globus1} T.Globus, M.Bykhovskaia, D.Woolard, and B.Gelmont
                     J.Phys.D:Appl.Phys. {\bf 36}, 1314 (2003).
   \bibitem{fabelinsky} I.L.Fabelinsky,  Molecular Light  Scattering,
                    Ch. VI (Moscow: Nauka) (1965).
   \bibitem{star}  E.V.Starikov, Physics Reports  {\bf 284}, 1 (1997).
   \bibitem{edwards} G.S.Edwards and Changle Liu, Phys.Rev.,
                     {\bf A44},  2709 (1991).
   \bibitem{zhang} C.T.Zhang, Phys.Rev. {\bf A40}, 2148 (1989).
   \bibitem{adair} R.K.Adair, Biophys.J. {\bf 82}  1147 (2002).
   \bibitem{chaplin} M.Chaplin,  Biophys.Chem. {\bf 83}, 211 (2000).
   \bibitem{mull} A.M\"{u}ller, H.B\"{o}gge and E.Diemann,
                    Inorg.Chem.Commun. {\bf 6}, 52 (2003).
   \bibitem{go} N.Go T.Noguti and T.Nishikawa,
                    Proc.Nat.Acad.Sci.USA, {\bf  80},   3696(1983).
   \bibitem{zakrz} T.H.Duong and K.Zakrzewska, J.Comp.Che.,
                   {\bf 18},  796 (1997).
   \bibitem{moliveanu} L.Movileanu, J.M.Benevides and G.J.Thomas Jr.,
                   Nucl. Acids Res., {\bf 30}, 3767 (2002).
   \bibitem{topal-fresco} M.D.Topal and J.R.Fresco,
            Nature {\bf 263}, 285 (1976).
   \bibitem{Auer}  Ch.Auerbach,
                   Mutation Research, London, John Wiley and Sons, (1976).
   \bibitem{Topal} V.I.Poltev; M.V.Kosevic,  V.S.Shelkovskii,
                V.A.Pashinskaja, E.X.Gonsales, A.V.Tepluxin, and
                G.G.Malenkov,
                Mol.Biology (Russian), {\bf 29}, 376 (1995).
   \bibitem{Robinson} H.Robinson, Yi-Gui Gao, Cornelia Bauer,
            Christopher Roberts, Christopher Switzer,
            H.-J.Andrew Wang,
            Biochemistry {\bf 37}, 10897 (1998).
        \bibitem{K1} H.Kamiya, H.Kasai, FEBS Lett.,
              {\bf 391}, 113 (1996).
        \bibitem{K2} H.Kamiya and H.Kasai,
             Biochemistry, {\bf 36}, 11125 (1997).
        \bibitem{K3} H.Kamiya and H.Kasai,
             J.Biol.Chem., {\bf 270}, 2595 (1995).
        \bibitem{S1} C.Switzer, S.E.Moroney, and S.A.Benner,
                   J.Am.Chem.Soc.Soc., {\bf 111}, 8322 (1989).
        \bibitem{suen} Wu Suen; T.G.Spiro, L.C.Sowers,
            J.R.Fresco,
            Proc.Natl.Acad.Sci. USA, {\bf 96}, 4500 (1999).
        \bibitem{S2} C.Switzer, S.E.Moroney, and S.A.Benner,
             Biochemistry, {\bf32 }, 10489  (1993).
        \bibitem{Persico} R.Geracitano, F.Persico,
            Physiol.Chem. and Physics, {\bf 3}, 361 (1971).
        \bibitem{Sukhorukov} B.I.Sukhorukov, V.I.Poltev,
            L.A.Blumenfeld,
            Biophysics (Russian), vol.9, 266 (1964).
        \bibitem{Sukh} B.I.Sukhorukov,
            Biophysics (Russian) 7, 664 (1962).
        \bibitem{SSSM} B.I.Sukhorukov, G.B.Sukhorukov,
            L.I.Shabarchina, and  M.M.Montrel,
             Biophysica (Russian) 45, 40 (2000).
        \bibitem{Kyogoku} Y.Kyogoku, M.Tsuboi, T.Shimanouchi, and
            .Watanabe,
             Nature 189, 120 (1961).
        \bibitem{Chris} C.A.Hunter,
            J.Mol.Biol. {\bf 230}, 1025 (1993).
        \bibitem{HunSand} C.A.Hunter, J.K.M.Sanders,
            J.Amer.Chem.Soc., {\bf 112}, 5525 (1990).
        \bibitem{Holstein} T.Holstein,
            Ann.Phys., vol.8, 325 (1959).
        \bibitem{Davy} A.S.Davydov,
            Sov.Phys. Usp. {\bf 251}, 898 (1982).
        \bibitem{Davy2} A.S.Davydov,
             Solid State Theory, Chapter X,
            Moscow, Nauka (1976), (in Russian).
        \bibitem{Vol} M.V.Volkenstein,
            Biophysics (in Russian), Moscow, Nauka, (1981).
        \bibitem{Nelson} P.Nelson,
            Proc.Natl.Acad.Sci. USA {\bf 96}, 14342 (1999).
        \bibitem{ma} M.Wang, M.J.Schnitzer, H.Yin, R.Landick,
            J.Gelles, and S.Block,
            Science. {\bf 282}, 902  (1998).
        \bibitem{LiuWang} L.F.Liu, J.C.Wang,
            Proc.Natl.Acad.Sci. USA, {\bf 84}, 7024 (1987).
        \bibitem{LeviCr} C.Levinthal, H.Crane,
            Proc.Natl.Acad.Sci. USA, {\bf 42}, 436 (1956).
       \bibitem{Peyr} T.Dauxois, M.Peyrard, A.R.Bishop,
            Phys.Rev., {\bf E47}, R44  (1993).
        \bibitem{99:b} V.L.Golo, E.I.Kats, Yu.M.Yevdokimov,
            Pisma ZhETF,  {\bf 70},  766 (1999).
        \bibitem{Sch1} V.L.Golo, E.I.Kats, and  M.Peyrard,
            Pisma ZhETF, {\bf 73}, 225 (2001).
        \bibitem{gv} V.L.Golo and Yu.S.Volkov,
                Int.J.Mod.Phys.C {\bf 14}, 133 (2003).
        \bibitem{AScot2} A.C.Scott,
            Phys.Reports., {\bf 217}, 1 (1992).
        \bibitem{Bodor} N.Bodor, M.J.S.Dewar, A.J.Hagert,
            J.Amer.Che.Soc., {\bf 92}, 2929 (1970).
        \bibitem{olson}    B.D.Coleman, W.K.Olson, and D.Swigon,
                       J.Chem.Phys. {\bf 118}, 7127 (2003).
        \bibitem{slun1} C.Kharif, E.Pelinovsky, T.Talipova, and
                        A.Slunyaev,
                        JETP Letters, {\bf 73}, 170 (2001).
        \bibitem{slun2} E.Pelinovsky, T.Talipova, and C.Kharif,
            Physica D, {\bf 147}, 83 (2000).
        \bibitem{alprot} A.Protogenov,
            JETP Lett., {\bf73 }, 292 (2001).
        \bibitem{chir2} J. Langowski, G. Chirico, and U. Kapp, \
                                    Structural Biology,
                                    Proceedings of the 8-th conversation,
                                    State University of New York, Albany, NY 1993.
        \bibitem{hagerman} P.J.Hagerman, Annu.Rev.Biophys.Biophys.Chem.
                       {\bf 17}, 265 (1988).
        \bibitem{grothers} D.M.Grothers, J.Drak, J.D.Kahn, and S.D.Levene,
                       Methods Enzymol. {\bf 212}, 2 (1992).
\end{thebibliography}
\end{document}